\pdfoutput=1
\documentclass[11pt,fleqn]{article}
\usepackage{xcolor}
\usepackage[english]{babel}
\usepackage[utf8]{inputenc}
\usepackage[T1]{fontenc}
\usepackage{lmodern}
\usepackage{amssymb}
\usepackage{mathtools}
\usepackage[lmargin=1in,rmargin=1in,bottom=1in,top=1in,twoside=False]{geometry}
\usepackage{enumitem}
\usepackage{amsfonts}
\usepackage{amsmath}
\usepackage{amsthm}
\usepackage{complexity}
\usepackage{mathrsfs}
\usepackage{graphics}
\usepackage{amsfonts,amssymb} 
\usepackage{url}              
\usepackage[pagewise]{lineno}
\usepackage{microtype}
\usepackage[defaultlines=4,all]{nowidow}
\usepackage[many]{tcolorbox}
\tcbuselibrary{theorems}
\usepackage[colorinlistoftodos,prependcaption,textsize=scriptsize]{todonotes}
\usepackage{xspace}
\usepackage{makecell}
\usepackage{pifont}
\usepackage{bm}
\usepackage{scalerel}
\usepackage{tabularx}
\usepackage{afterpage}
\usepackage[hidelinks]{hyperref}
\usepackage[nameinlink]{cleveref}
\usepackage{thm-restate}
\usepackage{float}
\usepackage{xspace}
\usepackage{subcaption}
\usepackage{algorithm}
\usepackage{algpseudocode} 
\usepackage{graphicx}
\usepackage{booktabs}
\usepackage{longtable}
\usepackage{multirow}

\algtext*{EndIf}
\algtext*{EndFor}
\algtext*{EndWhile}

\definecolor{myPurple}{RGB}{230,230,255}
\definecolor{interesting}{RGB}{140,0,60}
\definecolor{brown}{RGB}{210,180,140}

\renewcommand{\epsilon}{\varepsilon}
\renewcommand{\phi}{\varphi}

\newtheorem{theorem}{Theorem}[section]

\newtheorem{proposition}{Proposition}[section]

\Crefname{corollary}{Corollary}{Corollaries}
\Crefname{lemma}{Lemma}{Lemmas}
\Crefname{section}{Section}{Sections}

\newtheorem*{result*}{}
\newtheorem*{remark*}{Remark}

\usepackage{csquotes}
\usepackage[
    backend=biber,
    style=nature,
    sorting=none,
    maxbibnames=5,
    minbibnames=1
]{biblatex}

\begin{document}

\title{
A Hierarchy of Fast and Strong Signatures for Vertex Orbits}
\author{Joe Germany$^{1,2}$, Mohamad Al Sayed$^{2}$, Rana Dbouk$^{2,3}$, Amer E. Mouawad*$^{2,3}$ and Sara Najem*$^{2,4}$  \\ $^1$Department of Mathematics,  American University of Beirut, Lebanon \\ $^2$Complexity and Network Science Cluster, CAMS, American University of Beirut, Lebanon 
 \\ $^3$Department of Computer Science, American University of Beirut, Lebanon \\ 
$^4$Department of Physics,  American University of Beirut, Lebanon \\aa368@aub.edu.lb \\ sn62@aub.edu.lb  }
\date{}
\maketitle

\begin{abstract}
The defining features of a physical system are its symmetries and its invariants. For a network, the  symmetries are the automorphisms, and the invariants are the vertex orbits they define.
Two vertices of a graph may appear structurally indistinguishable without being interchangeable under any graph automorphism. 
Distinguishing apparent similarity from genuine symmetry lies at the heart of the vertex-orbit partition problem. In this work, we develop a hierarchy of increasingly expressive, automorphism-invariant signatures that draw on the spectral properties of the graph Laplacian and the Dirac operator to reveal subtle structural differences. Guided by a sequence of counterexamples, we progressively strengthen these signatures, ultimately incorporating probing techniques, which are temporary structural perturbations, that expose the hidden distinctions in the original unperturbed graph. We evaluate our approach across diverse benchmark families, recovering exact orbit partitions in challenging instances, including graphs that resist low-dimensional Weisfeiler--Leman refinement. We also examine the computational and numerical limitations of our constructions, distinguishing empirical success from theoretical completeness. Our findings highlight the potential of spectral and probing-based methods for uncovering the structural distinctions that define vertex orbit.
\end{abstract}


\section{Introduction}
The study of symmetries, or transformations that leave a system invariant, is central to physics \cite{landau1936theory,chaikin1995principles,goldenfeld2018lectures}. In statistical physics, the symmetries are the organizing principle behind collective behavior: they are used to classify phases of matter, constrain the structure of the order parameter and the free energy that governs it, and determine the type of phase transition that may arise \cite{nishimori2011elements,stanley1973introduction,kardar2007statistical,domb2000phase,fisher1974renormalization,stanley1999scaling}. The study of symmetry is not restricted to the continuous  and is equally applicable to systems defined on discrete structures, such as lattices. Complex networks, however, represent a more general class of discrete structures that model interactions in complex systems: the vertices carry the degrees of freedom, the edges carry the interactions, and the network's structure shapes the collective behavior that emerges \cite{tang2022optimizing,pecora2014cluster}. It is natural to ask, then, whether the statistical-physics notion of symmetry extends to the setting of a network and whether its symmetries play the same organizing role that they do for a system in continuous space \cite{sanchez2020exploiting,bayani2024transition,battiston2026collective,xiao2008emergence,nishikawa2016network,macarthur2008symmetry}. A first step in that direction is to compare the nodes of a network in pairs.  Two vertices may look identical without being interchangeable \cite{tononi1999measures,macarthur2009spectral,ward2022micro}. They may have the same degree, share the same distribution of distances to other vertices, or exhibit indistinguishable responses to a dynamical process, yet no symmetry of the graph need exchange them. Making this distinction precise is the task of \emph{vertex-orbit identification}. For a graph $\mathcal{G}=(V,E)$, its automorphism group $\operatorname{Aut}(\mathcal{G})$ consists of the permutations of $V$ that preserve adjacency. The orbit of a vertex $u$ is
\[
  \operatorname{Orb}_{\mathcal{G}}(u)=\{\pi(u):\pi\in\operatorname{Aut}(\mathcal{G})\},
\]
and the orbits partition $V$ into classes of genuinely interchangeable vertices. This partition is closely related to the graph-isomorphism problem, but it asks for more than a comparison of two graphs: it asks which vertices of a single graph are equivalent under its own symmetries. A standard way to approach such questions is to compute inexpensive, isomorphism-invariant vertex descriptors and refine the classes they induce
\cite{mckay2014practical}. The difficulty is that equality of descriptors need not imply membership in the same orbit. Indeed, obtaining the orbit partition efficiently on arbitrary graphs remains a challenging problem, against the backdrop of the quasipolynomial-time upper bound for general graph isomorphism \cite{babai2016graph}.
Broad accounts of the algorithmic and complexity-theoretic landscape are provided by Grohe and Schweitzer \cite{groheSchweitzer2020graph} and Grohe and Neuen \cite{groheNeuen2021recent}; the latter also surveys developments following Babai's quasipolynomial-time isomorphism algorithm \cite{babai2016graph}. Practical solvers such as \textsc{nauty} and \textsc{Traces} use combinatorial invariants, partition refinement, and individualization to determine graph symmetries and construct canonical labelings \cite{mckay2014practical}. 

Among the refinement schemes these methods build on, the Weisfeiler--Leman (WL) hierarchy provides a natural starting point; its details are given in The $k$-dimensional Weisfeiler--Leman algorithm section. Its successive color refinements extract increasingly detailed combinatorial information, with higher-dimensional variants operating on tuples of vertices; surveys of its algorithmic and logical role can be found in \cite{kiefer2020weisfeiler,groheNeuen2021recent}. Since the resulting colors are invariant under automorphisms, the induced vertex classes contain entire orbits, but distinct orbits may remain indistinguishable. The Cai--F\"urer--Immerman construction establishes fundamental limitations of bounded-dimensional WL for distinguishing non-isomorphic graphs \cite{cai1992optimal}. Such graph-identification lower bounds should not be conflated with failures of orbit recovery on a given graph. 

A complementary approach extracts invariants from matrices associated with a graph. The spectral properties of graph Laplacians and the limits of spectral characterization are treated in the surveys of Merris \cite{merris1994laplacian} and van Dam and Haemers \cite{vandamHaemers2003which}. Although isomorphic graphs have the same Laplacian spectrum, the converse fails in general; at the vertex level, cospectrality is likewise weaker than equivalence under an automorphism \cite{kempton2020characterizing}. Matrix functions such as the heat kernel retain information beyond a list of eigenvalues and give access to vertex-dependent, multiscale responses. Heat-kernel signatures were developed as intrinsic descriptors in geometric shape analysis \cite{sun2009heat}, and Raviv, Kimmel, and Bruckstein directly investigated graph isomorphisms and automorphisms through Laplacian spectral signatures \cite{raviv2013graph}. 
Thus, the use of heat diffusion to study graph symmetry has important precedents.

The wider relationship between random walks, diffusion, and network structure is reviewed by Masuda, Porter, and Lambiotte \cite{masuda2017random}. More directly, Najem and Mouawad \cite{najem2026finding} used curvature-inspired coefficients fitted to heat-kernel returns, followed by refinement and probing, to construct and verify candidate graph isomorphisms. The present work departs from that fitting-based geometric interpretation and focuses instead on recovering the orbit partition from heat-kernel information itself. Our signatures incorporate full diffusion rows organized by graph distance, as well as complementary operators and graph transformations. In particular, Laplacian-based correlation functions have been considered in network coarse-graining \cite{loures2023laplaciancoarsegrainingcomplex}, while graph Dirac operators provide coupled vertex--edge dynamics \cite{Bianconi_2021}. We use these ingredients as sources of additional invariant information, together with line graphs and exact-distance graphs, rather than treating any one operator as a complete characterization of vertex symmetry.

Increasing the amount of information available is not, however, a cost-free remedy. Richer graph representations can make dense spectral computations prohibitively expensive, while even combinations of several signatures may leave vertices from different orbits in a common class. Rather than indefinitely enlarge the transformed graphs, we turn to \emph{probing}. Here a vertex is distinguished not only by a signature of the original graph, but by the collection of signatures obtained after systematically applying controlled, rooted graph modifications. By considering admissible modifications in a way that respects relabeling, these probes expose relational differences that static summaries had concealed.

The outcome is a hierarchy of increasingly expressive signatures, not a claim that any one of them is a complete invariant for arbitrary graphs. We give the constructions, explain their automorphism-invariance in the exact setting, and document the counterexamples that motivated their successive refinements. We then evaluate the resulting partitions against known vertex orbits on a broad range of benchmark graphs, compare their distinguishing behavior with Weisfeiler--Leman refinement, and examine the time and memory costs of the spectral and probing stages. Under the stated arithmetic model, fixed-depth transformations and probes of fixed size admit polynomial bounds, but their degrees and practical costs can be substantial; moreover, numerical discretization introduces resolution limits absent from the exact mathematical objects. These distinctions matter: successfully recovering the orbit partition on difficult instances is evidence of discriminating power, not a general proof of completeness or of a polynomial-time orbit algorithm.

In this sense, the paper is both a continuation of and a departure from \cite{najem2026finding}. We retain the  central insight that diffusion can reveal subtle structural information, but no longer require that information to be interpreted as curvature on a latent manifold. Instead, we ask a more directly combinatorial question: \emph{which invariant observations of a vertex and which invariant responses to controlled changes in the graph are sufficient to distinguish it from vertices outside its orbit?} The counterexamples encountered along the way are not incidental obstacles; they are the organizing principle of the construction and the guide to its remaining limitations.

\section{Construction and evolution of our signatures}





We start from where Najem and Mouawad \cite{najem2026finding} left off. In that paper, the signatures are motivated by a direct analogy to a result for the heat kernel on a continuous, Riemannian manifold setting. In the short-time asymptotic, one can expand the heat kernel as an infinite series whose coefficients carry curvature information. More precisely, the classical Minakshisundaram–Pleijel expansion gives that the heat kernel obeys
\begin{equation}
    \label{eq:kernel_manifold}
    K(x,y,t) = \frac{1}{(4\pi t)^{n/2}} \, \exp \left( - \frac{d(x,y)^2}{4t} \right) \sum_{k=0}^\infty u_k(x,y) \, t^k,
\end{equation}
for an $n$-dimensional smooth Riemannian manifold, locally near the diagonal, as $t \rightarrow 0^+$. 
In particular, on the diagonal (when $x=y$), $u_0(x,x) = 1$, $u_1(x,x) = \frac{1}{6} R_\text{Scal}(x)$, and
\begin{equation}
    u_2(x,x) = \frac{1}{360} \left( 12 \Delta R_\text{Scal}(x) + 5 R_\text{Scal}(x)^2 - 2 |\text{Ric}(x)|^2 + 2 |\text{R}(x)|^2 \right),
\end{equation}
with
\begin{equation}\begin{aligned}
    |\text{Ric}(x)|^2 & \coloneqq \text{Ric}_{ij}(x) \, \text{Ric}_{ij}(x) \, g^{ii}(x) \, g^{jj} (x),  \\
    |\text{R}(x)|^2 &\coloneqq \text{R}_{ijkl}(x) \, \text{R}_{ijkl} (x) \, g^{ii}(x) \, g^{jj}(x) \, g^{kk} \, (x) \, g^{ll}(x),
\end{aligned}\end{equation}
where $g$ is the Riemannian metric, $R_\text{Scal}(x)$ is the scalar curvature at $x$, $\text{Ric}_{ij}(x)$ is the Ricci tensor, and $\text{R}_{ijkl}(x)$ is the Riemann curvature tensor. Higher-order terms are polynomials depending on the curvature tensor and its covariant derivatives.
For the mathematical literature, see \cite{chavel1984eigenvalues}, \cite[p.~201, Theorem 3.3.1]{gilkey2003asymptotic}, \cite{Minakshisundaram_Pleijel_1949}, and \cite[Theorem 3.1]{Suleymanova2017On}.

Conversly, a diffusion process on a graph $\mathcal{G}$ with $N$ nodes is given by the differential equation
\begin{equation}
   \frac{\partial u}{\partial t} = -L u, 
    \label{DiffusionOverG}
\end{equation}
where $u(t)$ is an $N$-vector representing a concentration variable over each node and $L$ is the Laplacian matrix. The latter is given by $L = D - A$, where $D$ is a diagonal matrix of the nodes' degrees and $A$ is the adjacency matrix satisfying $A_{ij} = 1$ if $(i,j) \in E$ and $0$ otherwise. Equation \ref{DiffusionOverG} admits the solution
\begin{equation}
    u(t) = e^{-Lt} \, u(0),
    \label{SolutionWithK}
\end{equation}
where $K = e^{-Lt}$ is the heat kernel of $\mathcal{G}$, which will take center stage in the investigations below, and $u(0)$ is the initial condition.

The paper \cite{najem2026finding} constructs vertex-based signatures by fitting an ansatz of the form 
\begin{equation}
\label{eq:heat_kernel_expansion}
    K(i, i, t) \approx \frac{1}{t^{d_s/2}} \sum_{k=0}^N u_k(i) t^k,
\end{equation}
to the heat kernel on a finite interval $[0,T]$. Here, the matrix heat kernel at each time $t \in [0,T]$ is obtained by the eigendecomposition $K(t) = e^{-Lt} = V e^{-\Lambda t} V^{-1}$ where $\Lambda$ is a diagonal matrix of the eigenvalues of the Laplacian $\lambda_1, \lambda_2, \dots, \lambda_n$ and $V$ is the matrix whose columns are the eigenvectors of $L$. To each node $i$ the tuple of coefficients $(u_k(i) )_{k=1}^N$ is attributed to obtain a partition of the graph by grouping nodes based on (rounded) equality of these tuples. \\
This algorithm offers a vivid geometric picture of the nodes in a graph ``perturbing'' a latent space in which the graph is embedded (see Figure 1 of \cite{najem2026finding}). One might imagine that discovering curvature information of that latent manifold might help partition the graph into ``similarly behaving nodes", thus motivating the formulation \eqref{eq:heat_kernel_expansion}.
A natural question then is: do the above signatures match the true orbit partition of a graph? How much information can we extract from this intuitive picture?

The route to attempting to answer this question starts by noting the shortcomings of the above approach. First, assuming that an ansatz of the form \eqref{eq:heat_kernel_expansion} holds, the dimension of a graph $\mathcal{G}$ is not a particularly well-defined notion. In \cite{najem2026finding}, the authors make use of the spectral dimension $d_s$, as a drop-in substitute for the dimension of the manifold in the continuous setting. The spectral dimension is obtained by linear regressing on the $\log$-$\log$ graph of 
\begin{equation}
    \label{eq:spectralDimension}
    \rho(\lambda' < \lambda) \sim \lambda^{d_s/2},
\end{equation}
for $\lambda \to 0$. It is a particularly suitable notion for lattice, fractal, and power-law/scale-free graphs \cite{torres2020simplicial}, due to the structure of the small eigenvalues of their Laplacians. For other graph classes, the small eigenvalue regime might be sparse or might not obey a power law of the form \eqref{eq:spectralDimension}, which drives the subdivision process that \cite{najem2026finding} adopts, leading to a growth in graph size, which became prohibitive. Even then, the notion of the spectral dimension is not presently rigorous, especially as to what counts as ``$\lambda \to 0$'' and arbitrariness in the choice of the interval to fit on, and it appeared difficult to find a rigorous justification without developing a substantial theory.
More importantly, the ansatz \eqref{eq:heat_kernel_expansion} is not rigorously justified in the graph setting. One can notice that for $d_s > 0$ and $u_0(i) \not= 0$, the above expansion $\frac{1}{t^{d_s/2}} \sum_{k=0}^N u_k (i) t^k$ goes to $\pm \infty$ as $t \to 0$.
However, the graph kernel behaves differently: 
at $t=0$, 
\begin{equation}
    K(x,y,t) = 
    \begin{cases}
        1 & \text{if } x=y \\
        0 & \text{if } x \not= y
    \end{cases},
\end{equation}
which is not captured in the above expansion. Additionally, we remark on the difference in the evolution of the heat kernel on manifold and on a graph. In particular, the Minakshusundaram-Pleijel expansion gives that the short-time asymptotic on a manifold is
\begin{equation}
    \lim_{t \to 0^+} t \log K(x,y,t) = - \frac{d_g(x,y)^2}{2},
\end{equation}
whereas, on a graph setting, Keller, Lenz, M\"unch, Schmidt, \& Telcs \cite{keller2016note} and Steinerberger \cite{steinerberger2018varadhan} prove considerably different short-time behavior, captured by
\begin{equation}
    \lim_{t \to 0^+} \frac{\log K(x,y,t)}{\log t} = d(x,y).
\end{equation}
Further, the presence of the $\log t$ term in the denominator indicates a mismatch between the two equations and implies that the graph kernel should not include a time-dependent exponential factor.


All of the above shortcomings inform the move to using the diagonal values of the heat kernel directly as signatures, that is, we endow the node $i$ with the signature 
\begin{equation}
    \label{sig:basicest}
    \text{sig} (i) = \{ K(i, i, t) : t = 0, \Delta t, 2\Delta t, \dots, T \},
\end{equation}
for a chosen time step $\Delta t$ and suitable large end-time $T$. 
We treat the signature as a set (or a lexicographically sorted array), so as to have automorphism-invariant vertex signatures, removing the effect of labeling.

\begin{figure}[ht]
    \centering
    \includegraphics[width=1.1\linewidth]{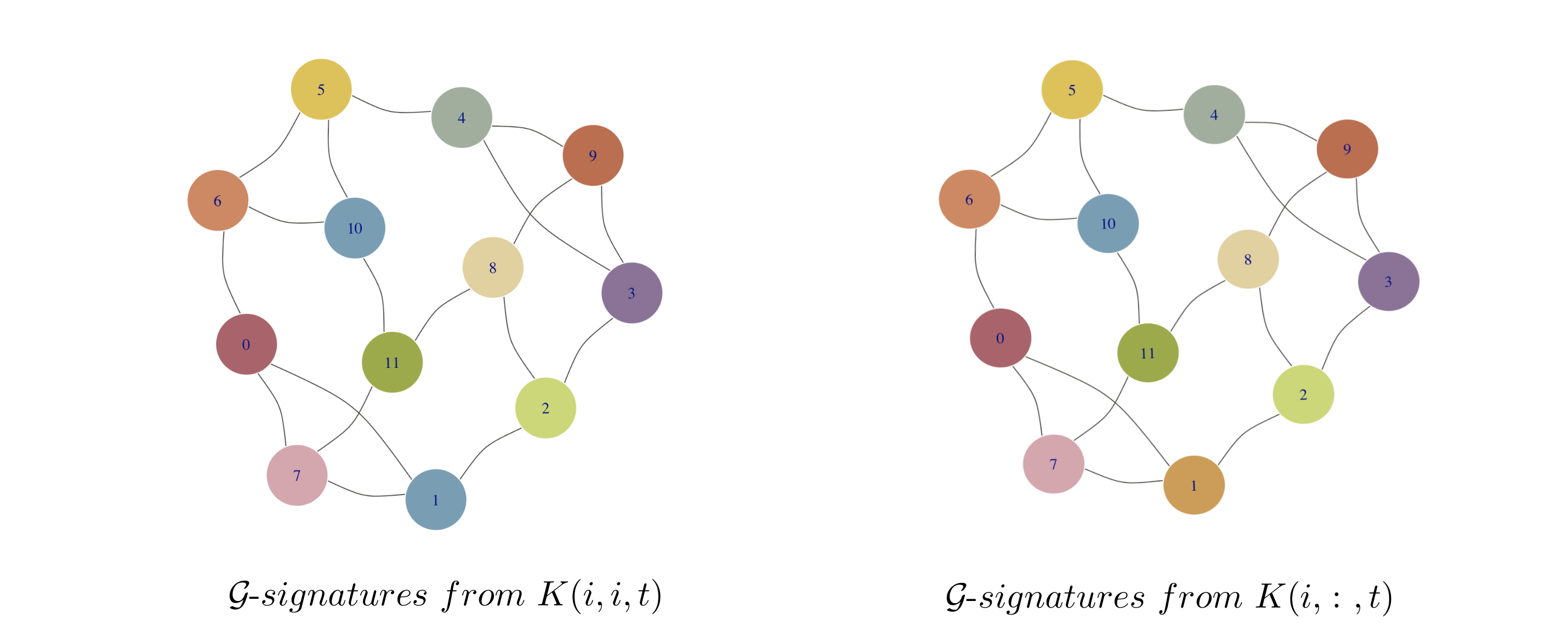}
    \caption{Frucht was not resolved using the diagonal of $K(i,i,t)$, which merged nodes 1 and 10 into one equivalence class, while the full rows $K(i,:,t)$ resolved the orbits and identified the nodes as singletons.}
    \label{fig:frucht}
\end{figure}

This is the concrete starting point of the investigation. The following discussion will aim to progressively find counterexamples to the proposed algorithms to obtain stronger and stronger signatures. First, signatures of the form \eqref{sig:basicest} do not distinguish the orbits of the Frucht graph, which possesses only a trivial automorphism (all nodes are in separate orbits), shown in Figure \ref{fig:frucht}. 
The above signatures merge nodes $1$ and $10$ into the same equivalence class, since they have the same heat kernel diagonal for all time. We prove this in Section ``Counterexamples'' of the Appendix in a computer-assisted Cayley-Hamilton-based argument. The mechanism of the proof follows from the observation that the heat kernel is represented mathematically by the matrix 
\begin{equation}
    e^{-Lt} = \sum_{k=0}^\infty \frac{(-1)^k t^k}{k!} L^k = I - tL + \frac{t^2}{2} L^2 - \frac{t^3}{3!} L^3 + \cdots.
\end{equation}
This infinite sum cannot be evaluated (computationally) without truncation or approximation.
However, by the Cayley-Hamilton theorem, every power $L^k$ with $k \geq n$ (the size of the matrix) can be expressed as a linear combination of $I, L, \dots L^{n-1}$, with coefficients depending on $k$ and $L$
Consequently, for each time $t$, the matrix exponential admits the exact representation
\begin{equation}
    e^{-Lt} = \sum_{j=0}^{n-1} \alpha_j (t) \, L^j,
\end{equation}
where the coefficients $\alpha_j(t)$ depend on $t$ and $L$.
The diagonal algorithm only captures the values $(e^{-Lt})_{ii}$, which represent solely the probability of a random walker that start at node $i$ to return to it at time $t$. We prove in the appendix that for the Frucht graph $(e^{-Lt})_{1,1} = (e^{-Lt})_{10,10}$, justifying their indistinguishability from the information only on the diagonal.

However, there is much more information in the heat kernel that is not captured in the diagonal, involving walks that start and end at different nodes. Najem and Mouawad \cite{najem2026finding} were cognizant of the presence of this information in the off-diagonal terms of the expansion. Nevertheless, they did not draw on it for the following reason: if the ansatz is adopted analogously with the Riemannian setting \eqref{eq:kernel_manifold}, then for $i\neq j$ the kernel has a dependence on $d$ through $\propto e^{-d(i,j)^2/4t}$. Then if the standard graph distance is used, 
and, for the reasons 
we delineated above, which
include subdivision, size growth, and a contrived spectral dimension, to list a few of them, it is not an adequate expansion for graphs. Otherwise, if the analogy with the latent manifold is pushed, with the latter's distance measure is unknown, then a nonlinear least-squares problem has to be solved that requires fitting for $u_k(i,j)$ and $d(i,j)$ simultaneously. The two parameter sets are coupled making their simultaneous estimation hard and ill-posed, in addition to being an evident source of error for a problem that is precision-critical. Hence, a way to circumvent the shortcoming of the diagonal and the fit using the off-diagonal elements is to shift to the full row element, in the form of the signatures
\begin{equation}
    \label{sig:fullRow}
    \text{sig} (i) = \{ K(i, j, t) : j =1, \dots, n; \ t = 0, \Delta t, 2\Delta t, \dots, T \}.
\end{equation}
Some information is lost in just looking at the values as a set (or as a lexicographically sorted array). We can retain more information by sorting the rows in a BFS-aware manner and preserving the temporal order, as such
\begin{equation}
    \label{eq:bfs_heat_kernel_sigs_at_one_time}
    \text{sig}_{\mathcal{G},\text{inter}} (i) (t) = \left( K(i,i,t) , \{K(i,j, t): d(i,j) = 1\}, \{K(i,j, t): d(i,j) = 2 \}, \dots  \right),
\end{equation}
and
\begin{equation}
    \text{sig}_{\mathcal{G},\text{inter}} (i) = \left( \text{sig}(0), \text{sig}(\Delta t), \text{sig}(2 \Delta t), \dots, \text{sig}(T) \right),
\end{equation}
with the final signature being
\begin{equation}
    \label{eq:full_sig_G}
    \text{sig}_{\mathcal{G}}(i) = \left( \text{sig}_\text{inter} (i) , \{ \text{sig}_\text{inter} (j) : d(i,j) = 1\}, \{ \text{sig}_\text{inter}(j) : d(i,j) = 2 \}, \dots  \right).
\end{equation}
We refer to the two levels of BFS as inner (Equation \ref{eq:bfs_heat_kernel_sigs_at_one_time}) and outer (Equation \ref{eq:full_sig_G}) BFS, referring to first ordering the kernel values based on distance from the source node, and then similarly for the full temporal row signature $\text{sig}_{\mathcal{G},\text{inter}}(i)$.
Loosely, this is motivated by a continuous analog: the Cartan-Ambrose-Hicks theorems says that if balls around points in two manifolds have matching curvature along corresponding geodesics, with directions compared by parallel transport, then those balls are isometric \cite[Lemma 1.41]{cheeger1975comparison}. Similarly, we imbue the same spirit in our signatures by grouping nodes based on distance to the source node $i$, as if batching nodes into concentric balls, making the signature comparison finer. With these signatures, the Frucht graph is resolved. 


Given the cost of this computation, further exacerbated by the need for multiple time steps (as shown in Complexity of the final algorithm Section), we seek cheaper alternatives to compute these signatures. To this end, we draw on the Laplacian-based approach to graph coarse-graining \cite{loures2023laplaciancoarsegrainingcomplex}. In this framework, correlations between nodes are computed using a statistical-mechanical model, and coarse-graining proceeds by merging the most strongly correlated pairs. Each node $i$ is assigned a scalar state variable $\phi_i$, whose interactions are described through the Hamiltonian
\begin{equation}
    \mathcal{H}(\Phi) = \frac{1}{2}\sum_{ij} \phi_{i}(L_{ij}+r\delta_{ij})\phi_{j} - \sum_{i} J_{i}\phi_{i},
\end{equation}
where $\Phi = (\phi_1, \dots, \phi_N)$ is the state of the network, $J_i$ is an external field applied to node $i$, and $r > 0$ is a regularization parameter that makes $L + rI$ invertible. The Laplacian term encodes interactions between adjacent nodes.
At inverse temperature $\beta$, the network states follow the Boltzmann distribution
\begin{equation}
    \mathcal{P_{\beta}}(\Phi) = \frac{e^{-\beta \mathcal{H}}}{\mathcal{Z}}, \qquad \qquad
    \mathcal{Z} = \int e^{-\beta \mathcal{H}(\Phi)} \, d\Phi,
\end{equation}
where $\mathcal{Z}$ is the partition function. The correlation between nodes $i$ and $j$ is their covariance
\begin{equation}
    C_{ij} = \overline{\phi_{j} \phi_{i}} - \overline{\phi_{j}} \, \overline{\phi_{i}},    
\end{equation}
where the overline denotes the average given by
\begin{equation}
    \overline{\phi_{i}} = \int \phi_{i} \,\mathcal{P_{\beta}}(\Phi) \, d\Phi,
\end{equation}
Since the Hamiltonian is quadratic, the Boltzmann distribution is Gaussian, with covariance matrix 
\begin{equation}
    C = \frac{1}{\beta} (L+rI)^{-1}.
\end{equation}
Setting $\beta = 1$, we obtain
\begin{equation}
    C = (L+rI)^{-1}.
    \label{thecorrMatrix}
\end{equation}

In what follows, we use the correlation matrix $C$ to build new signatures, similarly to Equations \eqref{eq:bfs_heat_kernel_sigs_at_one_time} and \eqref{eq:full_sig_G}, with
\begin{equation}
    \text{sig}_{C, \text{inter}} (i) = (C_{ii}, \{C_{ij} : d(i,j) = 1 \}, \{ C_{ij} : d(i,j) = 2 \}, \dots ), 
\end{equation}
and
\begin{equation}
    \label{eq:correlation_sig}
    \text{sig}_C (i) = (\text{sig}_{C, \text{inter}} (i), \{ \text{sig}_{C, \text{inter}} (j) : d(i,j) = 1 \}, \{ \text{sig}_{C, \text{inter}} (j) : d(i,j) = 2 \}, \dots).
\end{equation}
The correlation signatures are not an exact substitute for multiple time steps of the heat kernel; concretely, we prove that the signatures that use the heat kernel for all times are equivalent to the signatures that use the correlation matrix for all positive values of $r$. See Section ``Heat kernel and correlation equivalence'' in the Appendix for the proof.

\begin{figure}[ht]
    \includegraphics[width=1\linewidth]{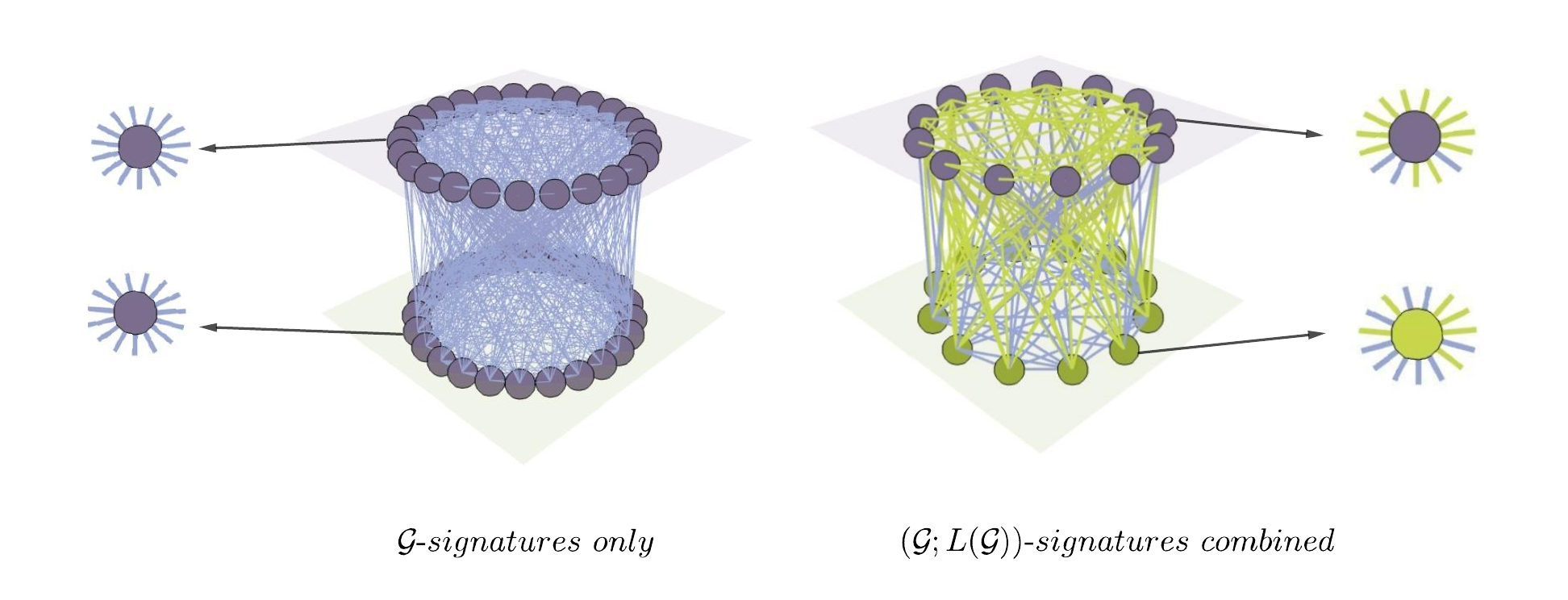}
    \caption{Nodes in orbit 1, lying in the lower plane, and those in orbit 2, lying in the upper plane get lumped into one big equivalence class using signature of the kernel of $\mathcal{G}$, and are subsequently undifferentiated using the node signature only. Using the information on the edges, computed from the kernel of $L(\mathcal{G})$, the signatures now can separate the equivalence class into two, which match the orbits.}
    \label{fig:sts13}
\end{figure}

However, in case of either the heat kernel \eqref{eq:full_sig_G} or the correlation matrix \eqref{eq:correlation_sig}, we run into another counterexample: the sts-13 graph from the Nauty dataset \cite{nautytracesgraphs}. The graph has two orbits, which get wrongly lumped into one equivalence class using our signatures, as shown in Figure \ref{fig:sts13}. The nodes are color-coded by their $\text{sig}(i)$, which clearly does not separate those of the first orbit, which lie in the lower plane, from those in the second orbit of the upper plane. See Section ``Counterexamples'' in the Appendix for a proof that the full heat kernel row signatures, with inner and outer BFS, do not distinguish these two nodes. Here we ask ourselves the question: are we making maximal use of the data in the adjacency? Our measurements so far focused on heat diffusing through the ``sensors'', which are the nodes, but did not consider edges, which are the other complementary building blocks of the graph that mediate diffusion. So, the natural question that follows is: what is happening to the heat across the edges? This required us to follow the diffusion on the line graph.

The line graph of a simple, undirected graph $\mathcal{G}=(V,E)$,
denoted by $L(\mathcal{G})=(E,E^1)$, is constructed by taking the
edges of $\mathcal{G}$ as the vertices of the line graph and
connecting two distinct vertices whenever their corresponding edges
share a vertex in the original graph. More precisely,
\begin{equation}
    \label{eq:lineGraph}
    (e,f) \in E^1 \iff e \neq f \quad \text{and} \quad e \cap f \neq \varnothing,
\end{equation}
where each edge is identified with the set of its endpoints. Iterated line graphs of $\mathcal{G}$ are defined recursively by 
by setting $L^0(\mathcal{G})=\mathcal{G}$
and $L^{k+1}(\mathcal{G})=L(L^k(\mathcal{G}))$. In particular, writing $L^k(\mathcal{G})=(V^k,E^k)$, the vertex set of each successive line graph is given by $V^{k+1}=E^k$, while its edges are determined by the same adjacency rule.

The signatures (whether heat kernel or correlation) are constructed by appending to the signature of each node $i$ the signatures of all the edges $(i,j)$ of the original graph that contain $i$, which are nodes in the line graph $L(G)$. This is represented mathematically as 
\begin{equation}
    \text{sig}_{(\mathcal{G},L(\mathcal{G})), \text{inter}} (i) = ( \text{sig}_{\mathcal{G}}(i), \{ \text{sig}_{L(\mathcal{G})}((i,j)) : (i,j) \in E \}),
\end{equation}
with the final signature being
\begin{equation}
\begin{aligned}
    \text{sig}_{(\mathcal{G}, L(\mathcal{G}))} (i) = (\text{sig}_{(\mathcal{G}, L(\mathcal{G})),\text{inter}} (i), \{ \text{sig}_{(\mathcal{G}, L(\mathcal{G})),\text{inter}} (j) : d(i,j)=1 \},& \\ 
    &\hspace{-5cm}\{ \text{sig}_{(\mathcal{G}, L(\mathcal{G})),\text{inter}} (j) : d(i,j)= 2 \}, \dots).
\end{aligned}
\end{equation}


% ...

\begin{figure}[ht]
    \centering
    
        \includegraphics[width=\linewidth]{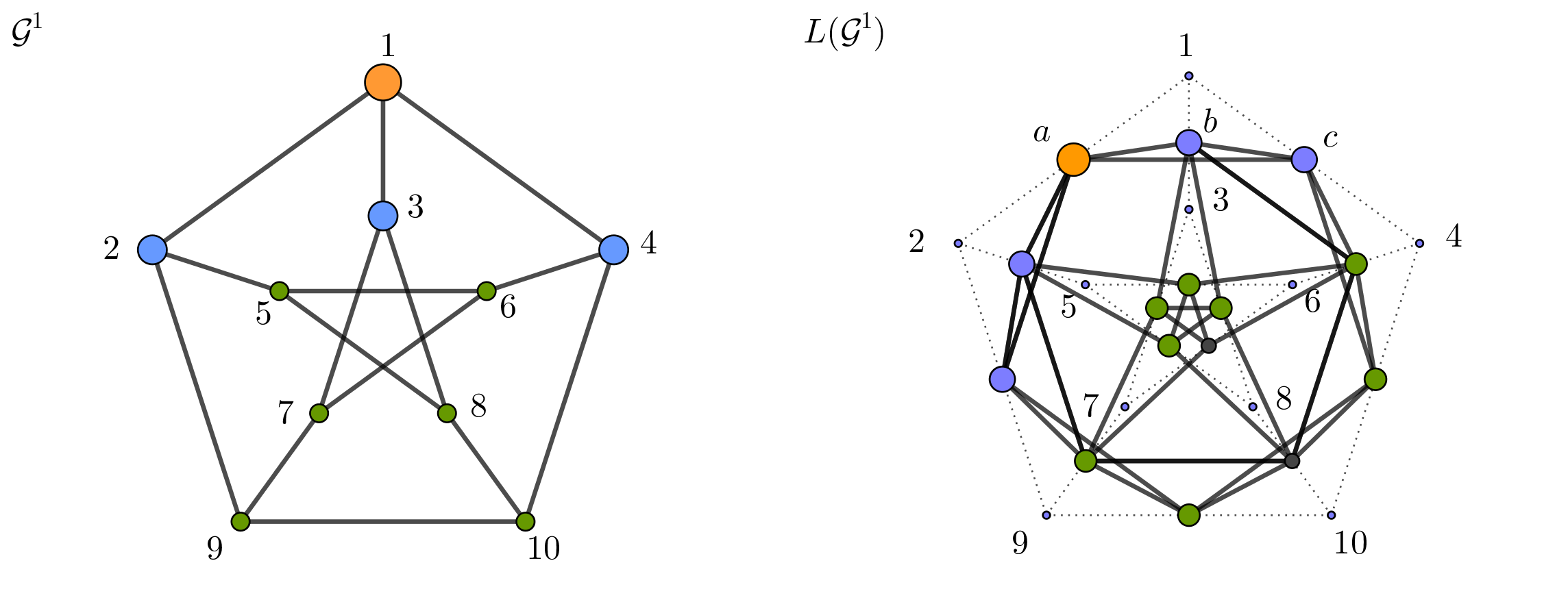}
       
    \caption{A sample graph shown in left  along with its line graph on the right. The orage is the node is the root, blue nodes belong to BFS level 1, green to level 2.  }
    \label{fig:renorm2}
\end{figure}

\begin{figure}[ht]
    \centering
    \includegraphics[width=0.9\linewidth]{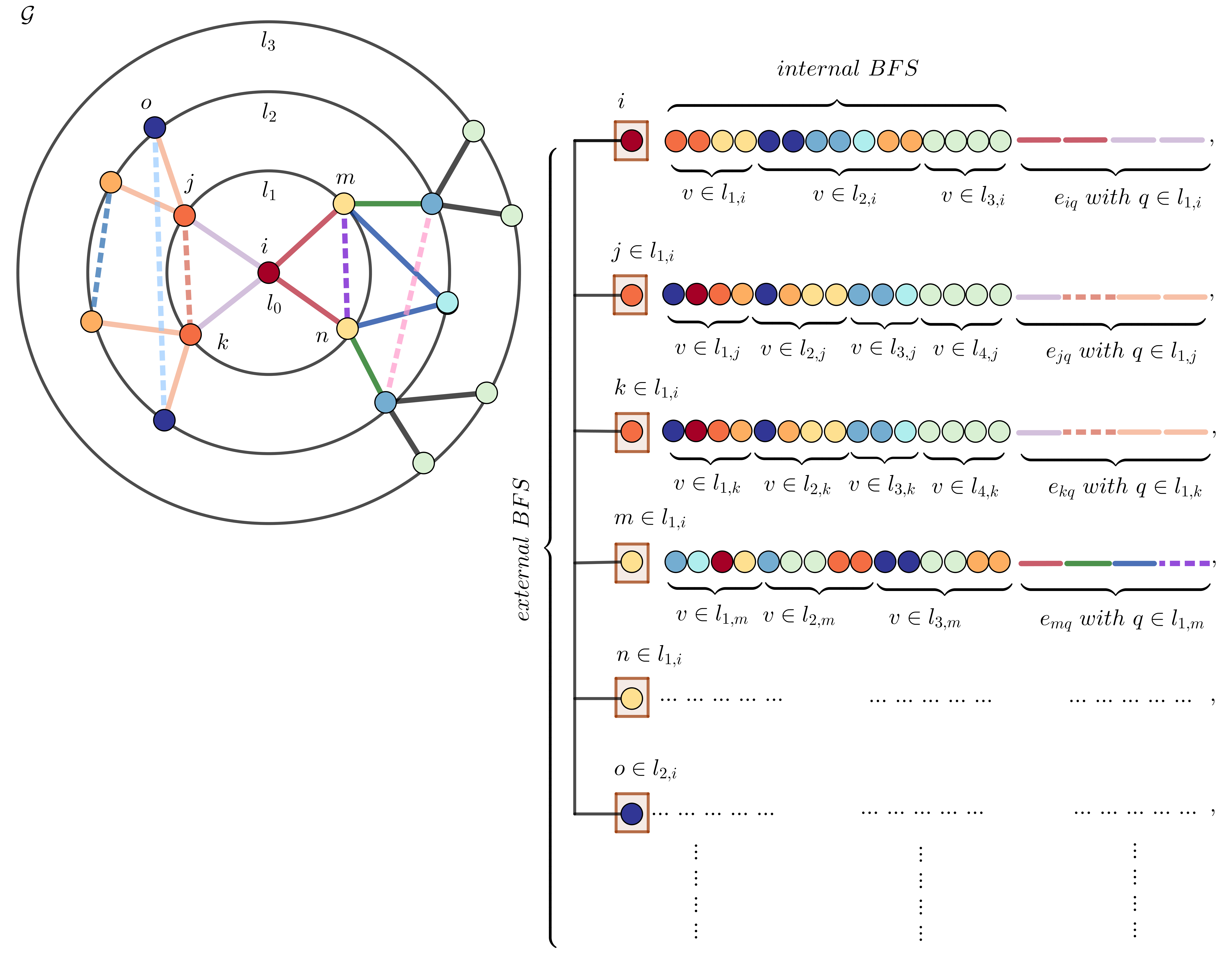}
    \caption{The updated signature data structure, which involves inner BFS followed by the edges incident to the node, inside an outer BFS.  }
    \label{fig:signbfsLg}
\end{figure}

In Figure \ref{fig:renorm2} on the left, a sample graph is shown with nodes color-coded and sized by their distance from root node 1, which is shown in orange. Blue is assigned to its first neighbors and green to its second. Similarly, its line graph is constructed on the right. The signature of node 1 contains its kernel values sorted by distance and appended to it the signatures of nodes $a$, $b$, and $c$, which are the edges incident to 1. The structure of the signature is now schematized in Figure \ref{fig:signbfsLg}.
These updated signatures lead to the resolution of sts-13, shown in the right pane of Figure \ref{fig:sts13}, where the nodes are now distinguished by the color of the edges incident to them. In particular, nodes of the first orbits have six green edges and nine violet, versus twelve green edges and three violet for nodes of the second orbit. 


Till now, the features incorporated in the nodes' signatures were mostly retrieved from the diffusion dynamics over the graph via its heat kernel. However, this dynamical process is a dissipative one; that is, the kernel decays exponentially in time, which describes information spread through the network. For this, we wanted to resort to a non-dissipative dynamical process. We propose the use of the Dirac operator \cite{Bianconi_2021}, which describes a wave dynamics on the graph $\mathcal{G}$. The advantage of this is two fold: first it offers an oscillatory behavior and thus non-decaying signatures of the nodes, and second, and most importantly, it couples to the dynamics of the nodes to those of the edges by construction. The Dirac operator for a graph $\mathcal{G}$ is an $(N+M)\times(N+M)$ matrix and is given by
\begin{equation}
    \mathcal{D} 
    =
    \begin{pmatrix}
    0 & B \\
    B^T & 0
    \end{pmatrix},    
\end{equation}
where  $B$ is the incidence matrix, an $N \times E$ matrix defined as
\begin{equation}
    B_{ij} =
    \begin{cases}
    0, & \text{if edge } j \text{ is not incident to node } i,\\
    1, & \text{if edge } j \text{ is directed towards node } i,\\
    -1, & \text{if edge } j \text{ is directed away from node } i.
    \end{cases}
    \label{definingB}    
\end{equation}
Clearly, the Dirac operator acts on both nodes and edges in $\mathcal{G}$, increasing the degrees of freedom associated with the dynamics from $N$, in diffusion processes, to $N+M$, in this case. 
We can then write the topological Dirac equation on $\mathcal{G}$, given by
\begin{equation}
    i\frac{\partial \varphi}{\partial t} = \mathcal{D} \varphi,
    \label{DefTopoDirac}
\end{equation}
The topological Dirac equation describes the time evolution of the state of the system encoded in the $(N+E)\times 1$ vector $\varphi$, defined over nodes and edges. The solution to Equation \eqref{DefTopoDirac} is
\begin{equation}
    \varphi = e^{-i\mathcal{D}t}\varphi(0).
\end{equation}
This evolution is analogous to the heat flow on $\mathcal{G}$ described by Equation~\ref{DiffusionOverG}, but the factor $i$ in the exponent produces oscillatory dynamics. To construct the signatures, we use the absolute values of the real and imaginary parts of each entry of $e^{-i\mathcal{D}t}$. Thus, for indices $x,y$ corresponding to nodes or edges, the microcomponents are the ordered pair
\begin{equation}
    \left(
        \left|\operatorname{Re}
            \left(e^{-i\mathcal{D}t}\right)_{xy}\right|,
        \left|\operatorname{Im}
            \left(e^{-i\mathcal{D}t}\right)_{xy}\right|
    \right).
\end{equation}
These absolute values remove the signs introduced by the choice of edge orientations.

At each time, we sort these pairs lexicographically, keeping the node and edge blocks separate. We then concatenate the results in time order and append to each node the sorted temporal signatures of its incident edges. The resulting intermediary signatures are used in the outer BFS construction.

\begin{figure}[ht]
    \centering
    \includegraphics[width=1\linewidth]{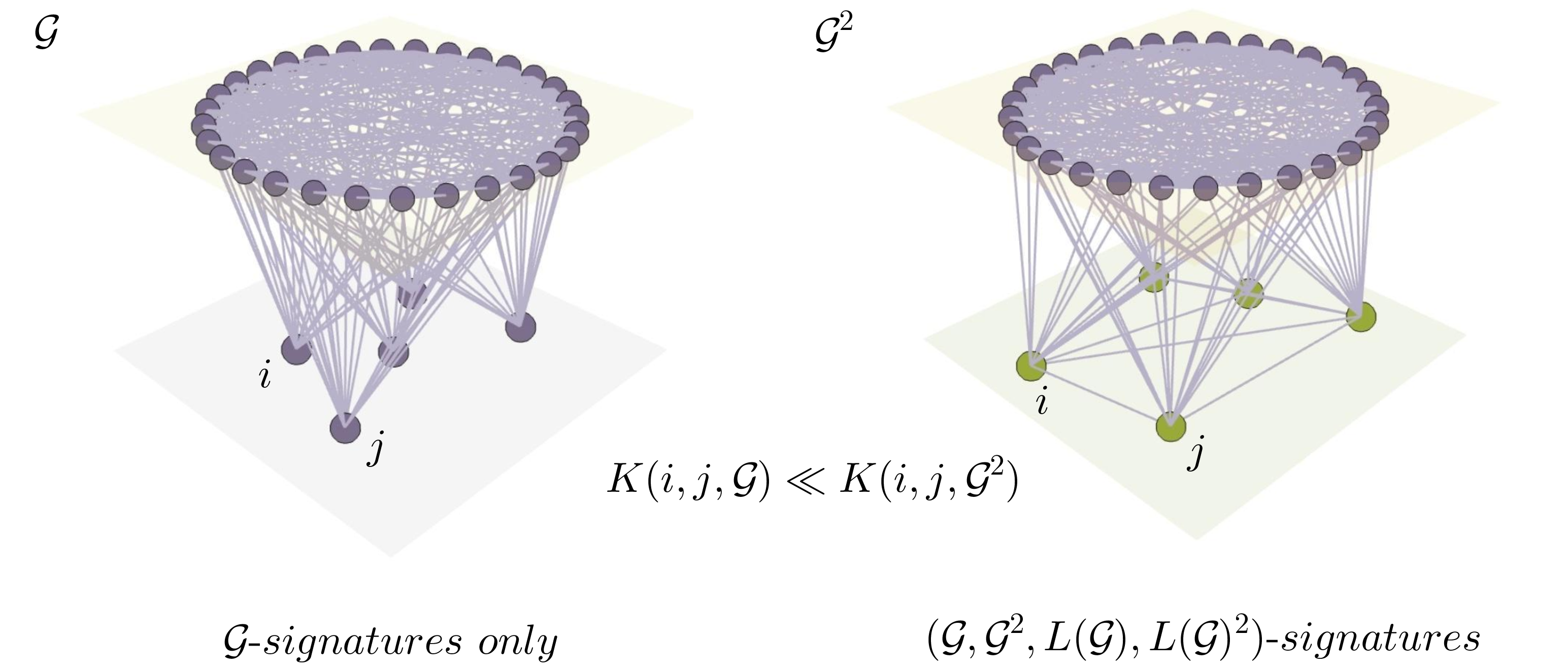}
    \caption{Nodes of orbit 1 do not communicate in $\mathcal{G}$, which are the ones that sit in the lower plane, while those in orbit 2 have intra links, and inter links that connect them to the first orbit. In $\mathcal{G}^2$ intra orbit 1 edges get created, which amplify their shared information, relative to what it was in $\mathcal{G}$.}
    \label{fig:sts15}
\end{figure}

\begin{figure}[ht]
    \centering
    \includegraphics[width=1\linewidth]{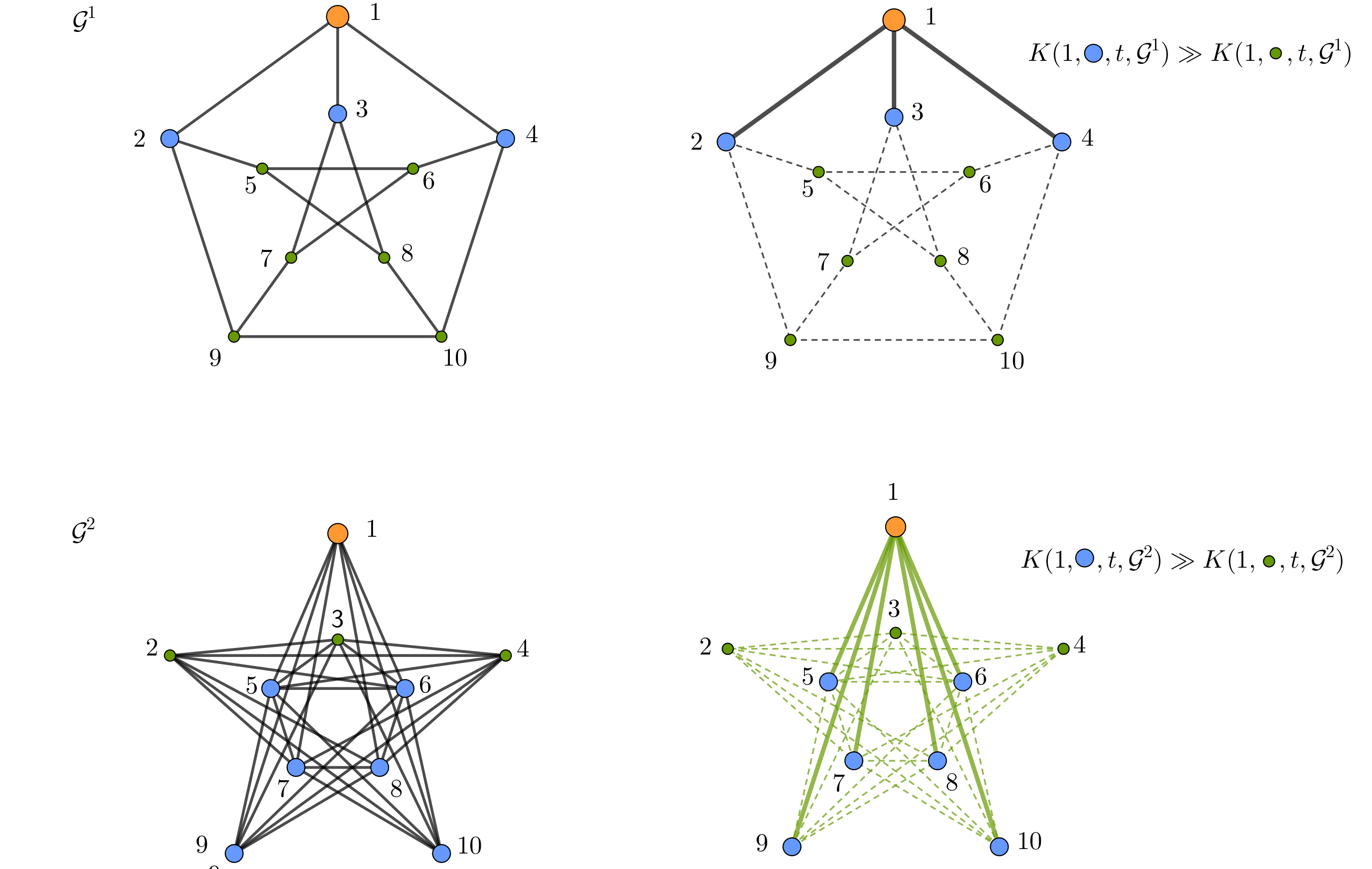}
        \caption{An illustration of information amplification, where the kernel values corresponding to the root node and its second neighbors in $\mathcal{G}$ get stronger on $\mathcal{G}^2$ which is constructed from the adjacency between nodes that are $d=2$ away in $\mathcal{G}^2$.  }
    \label{fig:renorm}
\end{figure}

At this stage, we thought we had a strong signature that could resolve all the sts graphs until we encountered sts-15, where the kernels and the correlations of the graph, its line graph, and the Dirac operator failed to resolve its orbits. A simple visualization of the orbits provided us with tremendous insight, shown in Figure \ref{fig:sts15}. The graph had two orbits also placed in the upper and lower planes, with one very noticeable feature: the nodes of the lower plane, which belong to the first orbit, were not connected. Since the kernel is a measure of the probability of visits, it was clear that the signal between nodes $i$ and $j$ of the first orbit was weak. In particular, given a root node shown in orange in Figure \ref{fig:renorm}, the elements of the kernel associated with its nearest neighbors, shown in blue, are stronger than those farther away, shown in green;   therefore, wanted to bring this small quantity to the front seat! 

Thus, to strengthen this structurally weak signal, we thought of a way to bring distant nodes closer and introduced the sequence of graphs $\mathcal{G}^{(k)}$. The latter is constructed using the nodes of $\mathcal{G}$ where a connection is made between two nodes in $\mathcal{G}^{(k)}$ if their pairwise shortest distance in $\mathcal{G}$ is $k$. Hence, the adjacency matrix of $\mathcal{G}^{(k)}$ is defined by
\begin{equation}
    A^{(k)}_{ij}
    =
    \begin{cases}
        1 & \text{ if $d (i,j) = k$ } \\
        0 & \text{ otherwise}
    \end{cases}.
\end{equation}
In Figure \ref{fig:renorm}, we show the relative kernel values for the nearest neighbors versus those of the second nearest neighbors in a graph $\mathcal{G}$. They switch order in $\mathcal{G}^2$. In particular, nodes 2,3, and 4, colored in blue, are the nearest neighbors of node 1 in $\mathcal{G}$. They become second neighbors in $\mathcal{G}^{(2)}$, and are subsequently colored in green, while nodes 5,6,7,8,9, and 10, which were second neighbors of 1 in $\mathcal{G}$ and were colored in green, now shift roles in $\mathcal{G}^2$ and become first neighbors and are thus colored in blue.  

\begin{figure}[ht]
    \centering
    \includegraphics[width=1\linewidth]{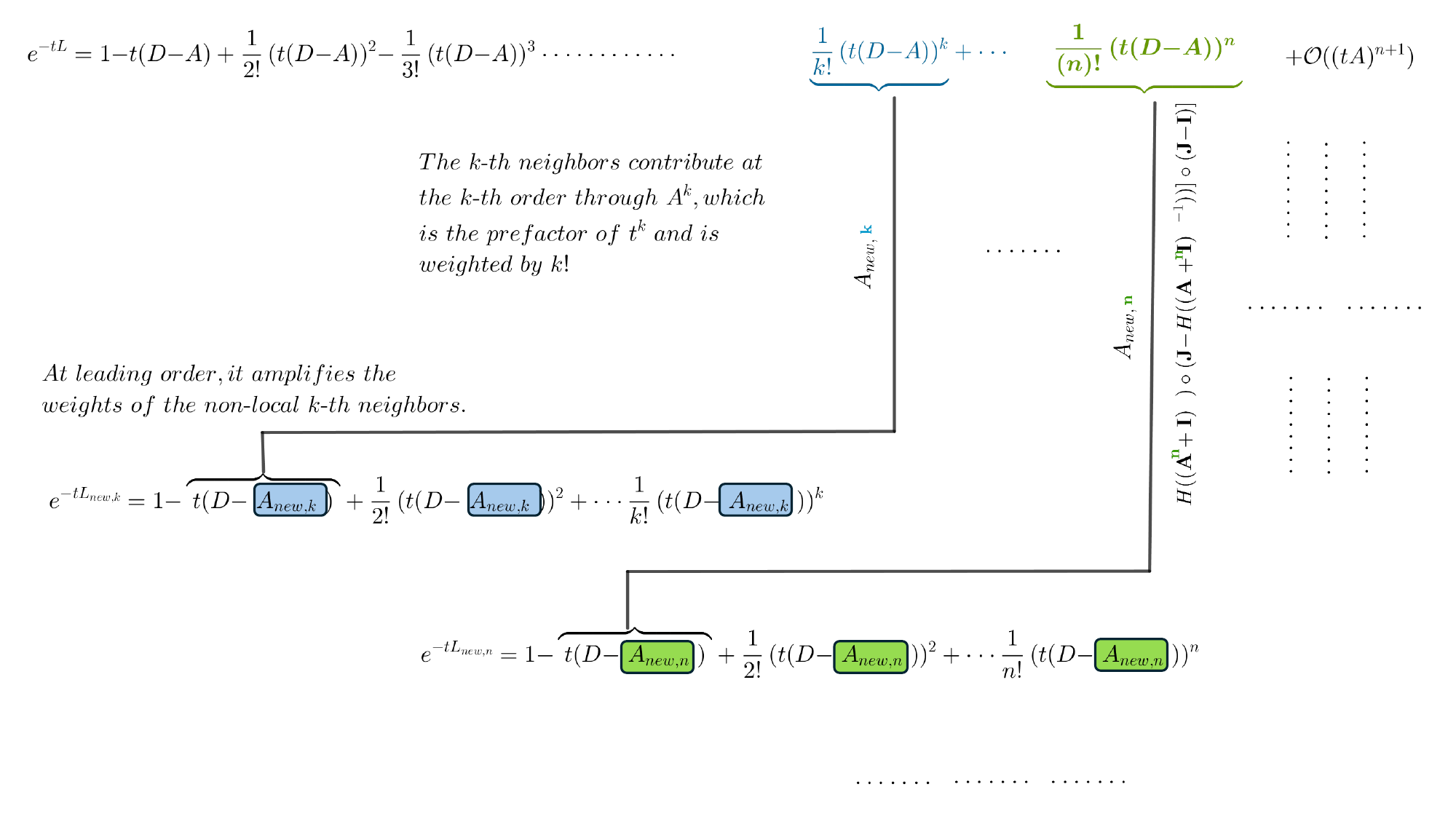}
    \caption{ The scheme shows that the terms which are $k\text{-}$distant from a node $i$ appear in the $k\text{-}$ element in the expansion, where their contribution is weakened by an inverse factorial of $k$.  $\mathcal{G}^{(k)}$ brings their contribution to the leading term, and thus as a renormalization procedure its acts an amplifier that effectively brings $\mathcal{O}(\epsilon)$ terms to $\mathcal{O}(1)$.  }
    \label{fig:perturb}
\end{figure}

\begin{figure}[ht]
    \centering
    \includegraphics[width=0.8\linewidth]{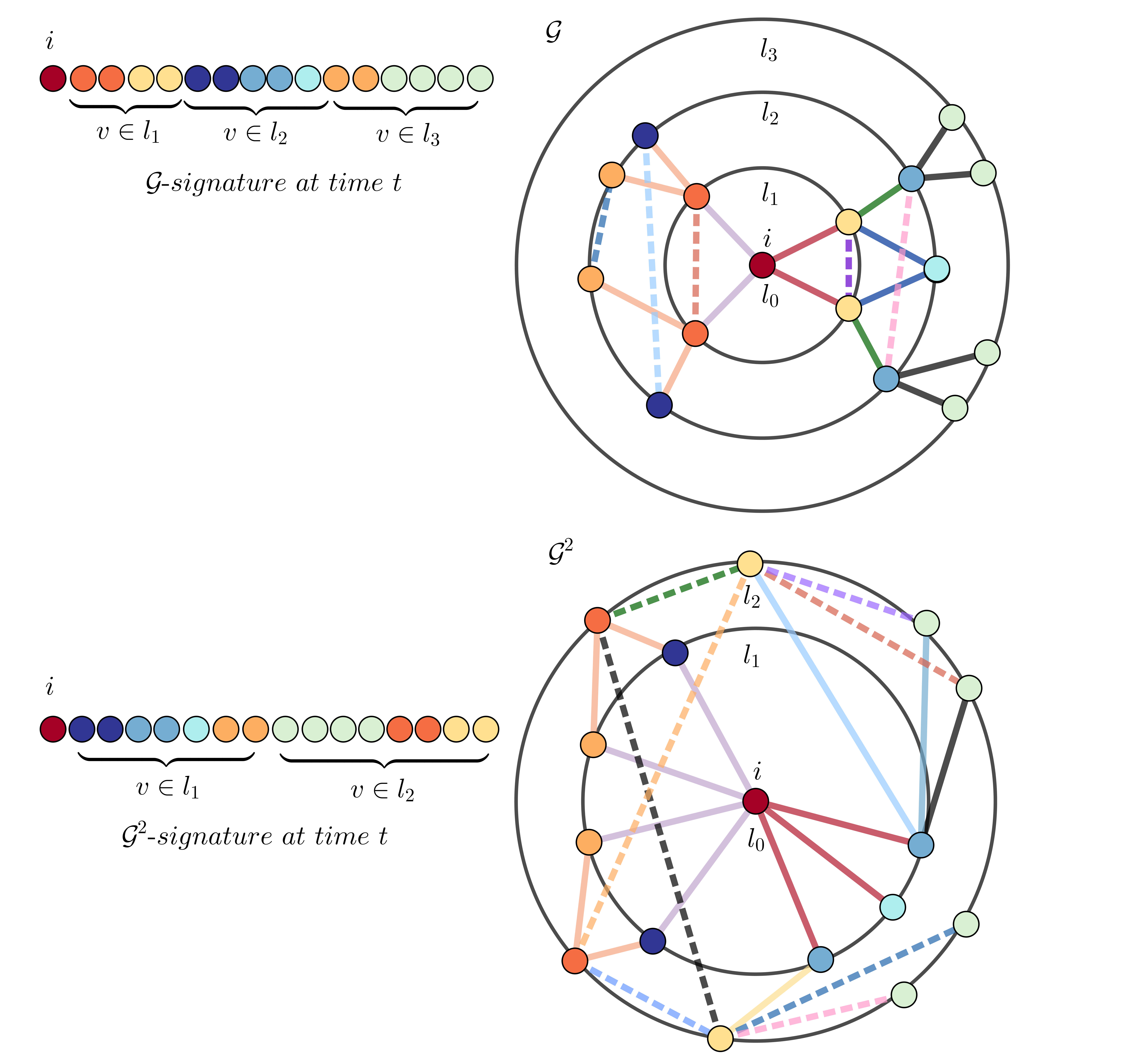}
    \caption{The colors of the nodes of $\mathcal{G}$ are shown together with $\mathcal{G}^2$, which is now added to the data structure of the signatures.  }
    \label{fig:signbfs}
\end{figure}

\clearpage
This approach bypasses intermediacy by making the connection between distant nodes immediate. In other words, it trades intermediacy with immediacy and subsequently brings information weakened by an inverse factorial weight between neighbors at distance $k$ in $\mathcal{G}$,  to the leading order as nearest neighbors in $\mathcal{G}^{(k)}$, which is illustrated in Figure \ref{fig:perturb}.
 The diffusion process is followed on these new structures using Equation \ref{DiffusionOverG}, which provides additional features for the nodes dictated by the behavior of the heat kernel of $\mathcal{G}^{(k)}$.
The updated signature includes those of $\mathcal{G}^d$, where $d$ here is the diameter of the matrix. An illustration is shown in Figure \ref{fig:signbfs}. 
This resolved the orbits of sts-15 which were differentiated by the signature of $\mathcal{G}$, $\mathcal{G}^2$, $L(\mathcal{G})$, and $L(\mathcal{G})^2$ since the graph has $d=2$, which is shown in the right pane of Figure \ref{fig:sts15}; it was also solved using the Dirac operator on $\mathcal{G}$ and $\mathcal{G}^{2}$ only. At this stage, we were cognizant of the prohibitive aspect of building the graph using $L^k(\mathcal{G})$, where $k$ is the line graph order, which grows like $(N^2)^k$. Two very hard graphs brought about a serious challenge that we had to face, for which a new strategy had to be devised.

In Section ``One direction of invariance'' of the Appendix, we prove that vertices in the same orbit of $\mathcal{G}$ receive equal signatures under all the constructions described above. Thus, unequal signatures certify that two vertices belong to different orbits, although equal signatures need not imply that they belong to the same orbit.

\section{The unyielding graphs and the probing that defeats them}
In this section, we report on the details of the two graphs which resisted our algorithm and had an uncompromising stiffness against any attempt to identify their orbits. These are had-52 (hadarmard-52 with $N = 208$), and pp-9-2 (projective plane with $N = 182$). 

\begin{figure}[ht]
    \centering
    \includegraphics[width=1\linewidth]{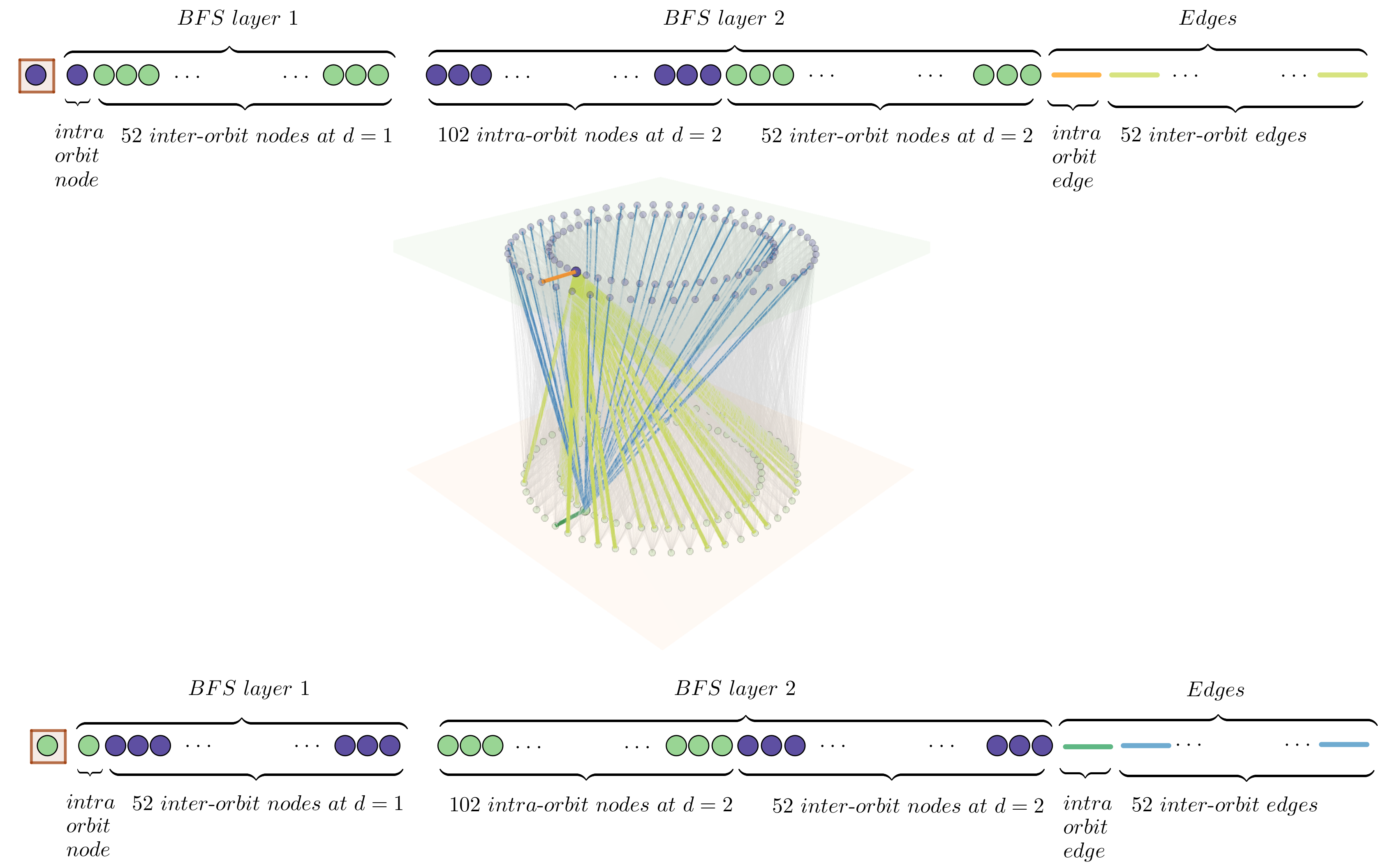}
    \caption{The real orbits of Hadamard-52 as they should be detected. The signatures fail to distinguish them and lumps them all in one orbit with similar node and edge signatures. }
    \label{fig:had52}
\end{figure}

The first, shown in Figure \ref{fig:had52}, has two orbits. Nodes that belong to the first orbit are colored in dark blue, while those that belong to the second are colored in green.  When a representative node from each orbit is considered as a root,  the colors of their BFS ordering  clearly distinguish them. In addition, the colors of their edges further serve as identifiers of the orbits. This is what one would expect the signature to capture; alas, it does not.  Our failing signature concatenated the kernels of $\mathcal{G}$, $\mathcal{G}^2$, $L(\mathcal{G})$, $L(\mathcal{G})^2$, the correlations, and the Dirac operators that incorrectly lumped all the nodes in one equivalence class.  Obviously, $L(L(\mathcal{G}))$ pushed our graph size to $\sim N^4$, prohibiting the computation of the kernel at a cost of $ \sim (N^4)^3 = N^{12}$. A different strategy to break it started to brew.


\begin{figure}[ht]
    \centering
    \includegraphics[width=0.5\linewidth]{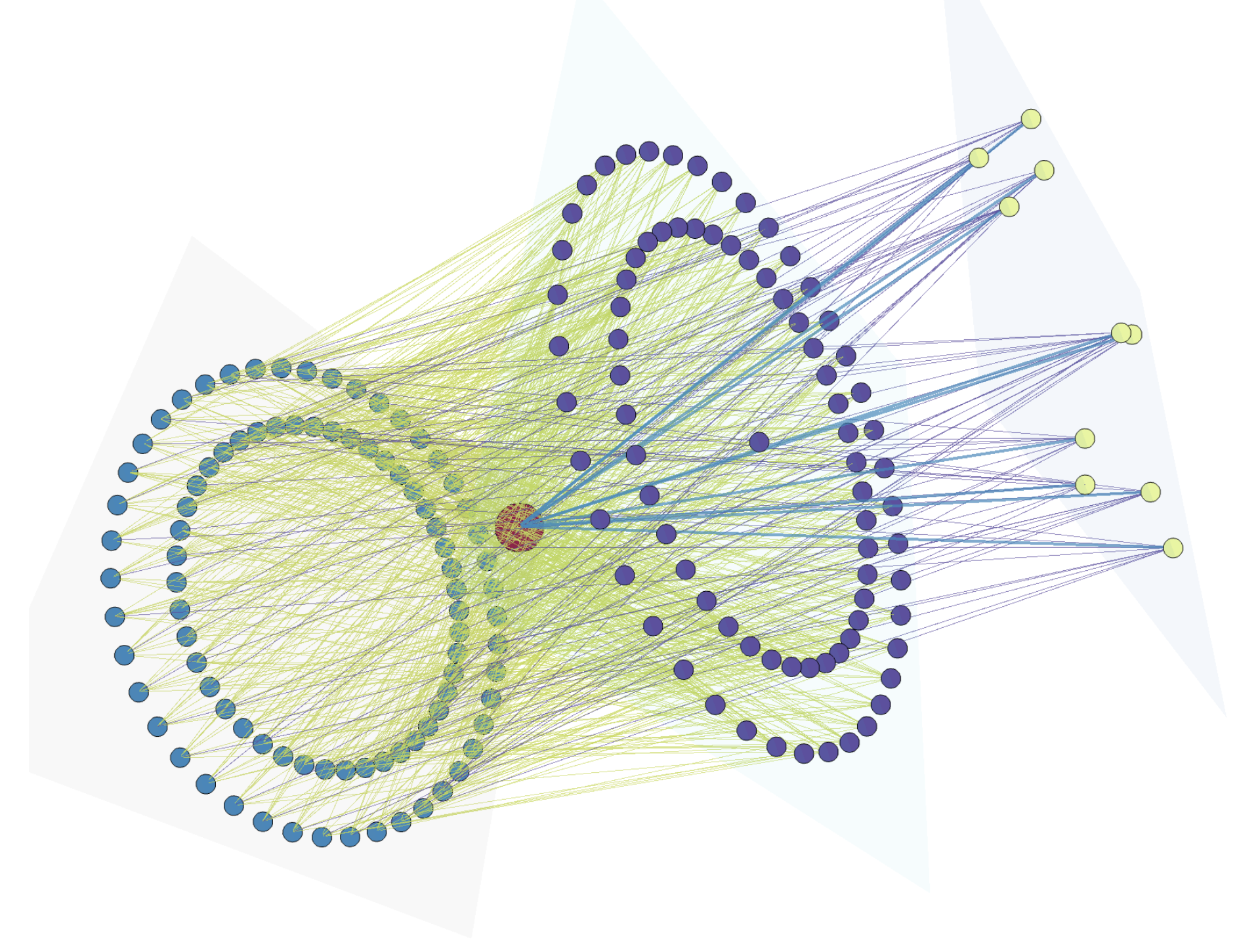}
    \caption{The orbits of pp-9-2 are shown in the figure lying in three planes along with a singleton orbit colored in dark orange.  }
    \label{fig:pp92}
\end{figure}

The second graph, pp-9-2, shown in Figure \ref{fig:pp92} has four orbits, a singleton, an orbit with 10 nodes, and two others. Again, it resisted our thus far invincible signatures, which in this case went to a depth of $d=3$, concatenating $\mathcal{G}$, $\mathcal{G}^2$, $\mathcal{G}^3$ $L(\mathcal{G})$, $L(\mathcal{G})^2$, $L(\mathcal{G})^3$, correlations, and the Dirac operators.  This failure, which also lumped all the nodes in the same equivalence class, provided a ferment for another orbit identification strategy.

Given these two very challenging graphs, we resorted to probing schemes. In particular, we describe the ones that hammered away these two graphs' resistance: the folding scheme and the neighborhood deletion. 

In order to distinguish vertices with identical signatures, we introduce probing techniques that apply temporary modifications to the graph and recompute signatures on the resulting graphs. These modifications aim to reveal structural differences that are not captured by the original signatures, allowing the separation of vertices belonging to distinct orbits.

Let $\mathcal{G}=(V,E)$ be a graph, and let $\tau_{\mathcal{G}}(u)$ denote the signature of a vertex $u\in V$. For each vertex $u$, a probing technique generates a collection of modified graphs $\mathcal{G}'_1,\mathcal{G}'_2,\ldots,\mathcal{G}'_m$. 
Each modified graph $\mathcal{G}'_i$ contributes a signature value $s_i(u)$ to the probing profile of $u$. Depending on the probing technique, these values may be derived from the signature of $u$ itself or from the signatures of other vertices in the modified graph. The probing profile is defined as
\begin{equation}
    P(u) = \operatorname{sort}\bigl(s_1(u),s_2(u),\ldots,s_m(u)\bigr),
    \label{eq:probing-profile}
\end{equation}
where sorting ensures that the profile is independent of the order in which the signatures are collected, while preserving repeated values. Probing is applied independently to each equivalence class of the current vertex partition. For each class, a probing profile is constructed for every vertex, and vertices are separated into new classes according to their profiles.

Our implementation supports several probing techniques. We describe below the techniques used in our experiments, each applied to a selected vertex $u\in V$. The remaining supported techniques are detailed in Appendix~\ref{app:probing}.

\begin{figure}[ht]
    \centering
    \includegraphics[width=1.\linewidth]{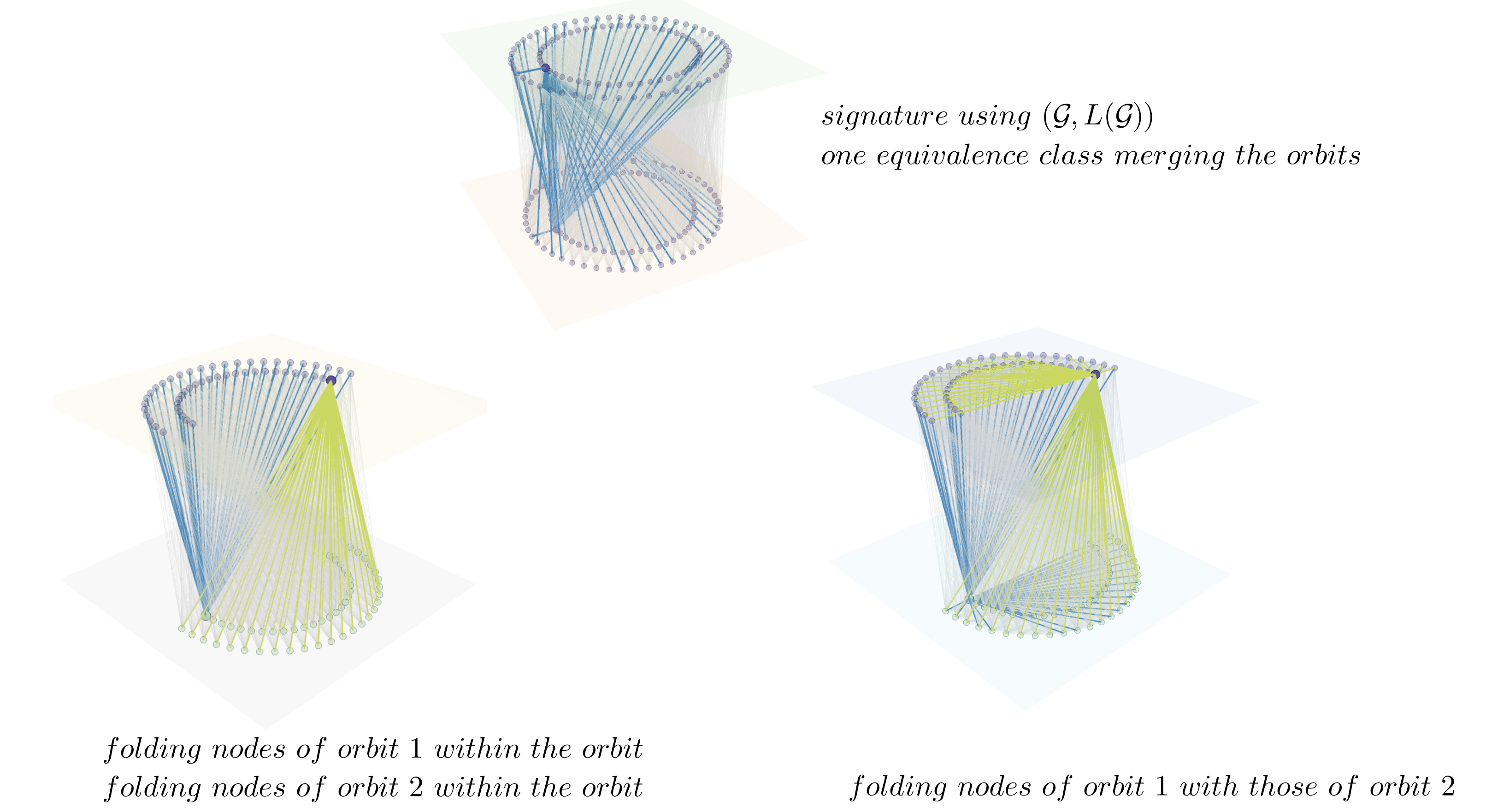}
    \caption{The figure is an illustration of the folding procedure. When the two-nodes folding probe is applied, the resulting signatures allow to distinguish the orbit. The colors of the edges incident to a node of the first orbit are different from those incident to a node in the second.  }
    \label{fig:had52folded}
\end{figure}

Folding selects $M\geq 1$ disjoint subsets $S_1,\ldots,S_M\subseteq V$, each containing $p\geq 2$ vertices, with $u\in S_1$. The vertices of each subset are merged into a single vertex, combining their incident edges and removing self-loops and duplicate edges. The subset containing $u$ is represented by $u$ in the modified graph. All possible selections of such disjoint subsets are considered, and the resulting signatures $\tau_{\mathcal{G}'}(u)$ are collected to construct $P(u)$. This strategy, for $p=2$, resolved the orbits of had-52, which is shown in Figure \ref{fig:had52folded}. Note that the nodes' signatures were indistinguishable before probing. The resulting signatures of folding two nodes of the same orbit, and two nodes of different orbits yield very distinct signatures, leading to the recovery of the correct colors. 

\begin{figure}[ht]
    \centering
    \includegraphics[width=0.8\linewidth]{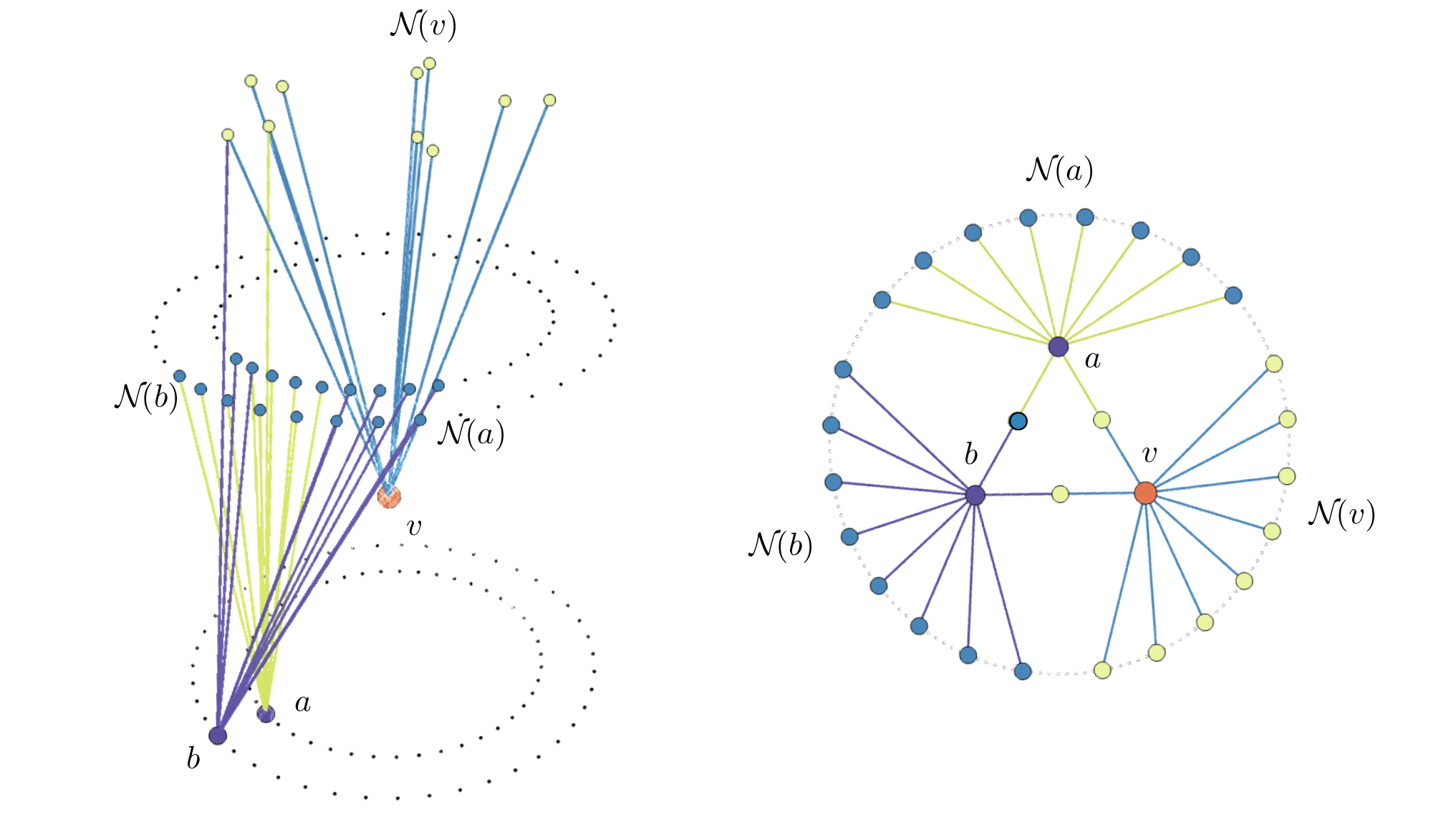}
    \caption{The neighborhood deletion probe is illustrated. The figure shows the correctly resolved orbits for pp-9-2, with three sample nodes, $a$, $b$, and $v$ along with their neighborhoods $\mathcal{N}(a)$, $\mathcal{N}(b)$, and $\mathcal{N}(v)$. The resulting deleted neighborhood is also shown.   }
    \label{fig:pp92probe}
\end{figure}

Neighborhood deletion selects a subset $S\subseteq V$ of $p\geq 2$ vertices containing $u$. Let
\[
    N[S]=\bigcup_{v\in S}(N(v)\cup\{v\}).
\]
Remove all vertices in $N[S]$, then reintroduce $u$ and connect it to every remaining vertex. An optional parameter $d_{\max}$ restricts the selected vertices to be pairwise within distance $d_{\max}$. The profile $P(u)$ collects $\tau_{\mathcal{G}'}(u)$ over all valid selections of $S$. This strategy resolved the orbits of pp-9-2 for $p=3$ and $d = 2$, with a sample deleted neighborhood shown in Figure \ref{fig:pp92probe}. 

\section{Complexity of the final algorithm}

Having introduced the individual components of our signature construction, we now combine them into a unified hierarchical refinement procedure for our experimental evaluation. The procedure incrementally strengthens the vertex signatures by incorporating spectral operators, graph transformations, and probing strategies, allowing us to assess the contribution of each component to refining the vertex partition. 
Algorithm~\ref{alg:hierarchy} summarizes this procedure, in which each stage preserves the distinctions established by the preceding stages \footnote{Code available at \url{https://github.com/joeGermany/graph_isomorphism_curvature}}.

\begin{algorithm}[t]
\caption{Hierarchical signature-based orbit refinement}
\label{alg:hierarchy}
\begin{algorithmic}[1]
\Require Graph $G=(V,E)$, maximum line-graph order $J$,
probing strategies $\mathcal{Q}$, true orbits $\mathcal{O}$
\Ensure Refined vertex partition $\Pi$

\State $\Pi \gets \{V\}$
\State $\mathcal{A} \gets
\{\text{Laplacian},\text{Correlation},\text{Dirac}\}$

\For{$j=0$ to $J$}
    \State $H \gets L^j(G)$
    \For{$d=1$ to $\operatorname{diameter}(H)$}
        \State $S \gets$ distance-$d$ graph of $H$

        \ForAll{$a \in \mathcal{A}$}
            \State $\sigma \gets$ Spectral signatures on $S$ using $a$
            \State $\Pi \gets \mathrm{Refine}(\Pi,\sigma)$
            \If{$\Pi = \mathcal{O}$}
                \State \Return $\Pi$
            \EndIf
        \EndFor

        \ForAll{$q \in \mathcal{Q}$}
            \State $\sigma \gets$ Correlation signatures on $S$
            using probing technique $q$
            \State $\Pi \gets \mathrm{Refine}(\Pi,\sigma)$
            \If{$\Pi = \mathcal{O}$}
                \State \Return $\Pi$
            \EndIf
        \EndFor
    \EndFor
\EndFor

\State \Return $\Pi$
\end{algorithmic}
\end{algorithm}

We analyze the worst-case time and space complexity of our approach by examining its main computational components. We first consider vertex signature construction using the Laplacian heat kernel, Dirac operator, and correlation matrix. We then examine how these operations scale across the original graph and its successive distance-based and line-graph transformations, including signature aggregation and partition refinement. Finally, we analyze the probing techniques and combine the results to derive the overall bounds for our algorithm.

Let $\mathcal{G}=(V,E)$ denote a graph processed by the algorithm, with $N=|V|$ vertices and $M=|E|$ edges. Let $T$ denote the number of sampled time points for time-dependent operators. We assume fixed-precision arithmetic and standard dense linear algebra, where eigendecomposition, matrix inversion, pseudoinversion, and multiplication of two dense matrices require $O(d^3)$ time and $O(d^2)$ space. We assume comparison-based sorting with $O(d\log d)$ worst-case time for $d$ scalar values. 
All-pairs shortest-path distances are computed using SciPy's Floyd--Warshall algorithm, requiring $O(N^3)$ time and $O(N^2)$ space.

For the Laplacian heat kernel, computing the eigendecomposition of its $N\times N$ matrix requires $O(N^3)$ time and $O(N^2)$ space. At each of the $T$ sampled time points, computing the full heat-kernel matrix rows requires $O(N^3)$ time. Computing all-pairs shortest-path distances requires $O(N^3)$ time, while sorting the heat-kernel matrix rows based on these distances requires $O(N^2\log N)$ time per time point. Quantizing the entries and refining the vertex equivalence classes require an additional $O(N^2\log N)$ time per time point. Since the remaining operations are dominated by the heat-kernel computations, the overall worst-case time complexity is $O(TN^3)$. The resulting signatures are then aggregated according to BFS layers. Performing BFS from every vertex requires $O(N(N+M))$ time. Since each raw signature has $O(TN)$ values, sorting these signatures within the BFS layers requires at most $O(TN^3\log N)$ time across all vertices. Therefore, the overall worst-case time complexity, including BFS aggregation, is
BFS aggregation, is
\[
O(TN^3\log N).
\]

For Laplacian space complexity, the implementation stores the heat-kernel values across all $T$ sampled time points, requiring $O(TN^2)$ space. The eigendecomposition, distance matrix, and intermediate matrices require an additional $O(N^2)$ space. BFS aggregation reuses the stored raw signatures rather than duplicating their values, requiring at most $O(N^2)$ additional space for the references across all vertices. Therefore, the overall worst-case space complexity is
\[
O(TN^2+N^2)=O(TN^2).
\]

The Dirac matrix has dimension $(N+M)\times(N+M)$, and computing its eigendecomposition requires $O((N+M)^3)$ time and $O((N+M)^2)$ space. At each of the $T$ sampled time points, constructing the full Dirac kernel requires $O((N+M)^3)$ time. To construct the vertex signatures, the kernel entries corresponding to each vertex are divided into two groups: $N$ vertex-to-vertex entries and $M$ vertex-to-edge entries. The quantized real and imaginary value pairs are sorted lexicographically within each group separately, and both sorted groups are retained as components of the vertex signature. Similarly, each edge signature is constructed by separately sorting its $N$ edge-to-vertex entries and $M$ edge-to-edge entries. These components are collected across all $T$ time points to form the initial vertex and edge signatures. Sorting the two groups for a single vertex or edge requires $O(N\log N+M\log M)$ time. Since there are $N$ vertices and $M$ edges, the total sorting time per time point is
\[
O\left((N+M)(N\log N+M\log M)\right)
=O((N+M)^2\log(N+M)).
\]

After processing all time points, each vertex collects and lexicographically sorts the complete signatures of its incident edges. For a vertex $v$, sorting its $\deg(v)$ edge signatures requires $O(\deg(v)\log\deg(v))$ comparisons. Since $\deg(v)\leq N-1$, this is bounded by $O(\deg(v)\log N)$. Summing over all vertices and using $\sum_{v\in V}\deg(v)=2M$, the total number of comparisons is
\[
O\left(\sum_{v\in V}\deg(v)\log N\right)
=O(M\log N).
\]
Each comparison may examine up to $O(T(N+M))$ values, and the overall sorting time is $O(TM(N+M)\log N)$. Since the sorting operations are bounded by the kernel computations, the overall time complexity of constructing the raw Dirac signatures is
\[
O(T(N+M)^3).
\]
The resulting signatures are then aggregated according to BFS layers. Performing BFS from every vertex requires $O(N(N+M))$ time. Each raw Dirac vertex signature contains its own kernel entries and the signatures of its incident edges, giving at most $O(TN(N+M))$ values. Comparing two such signatures may therefore require $O(TN(N+M))$ time. Since sorting the BFS layers requires at most $O(N^2\log N)$ comparisons across all vertices, the overall time complexity, including BFS aggregation, is
\[
\begin{aligned}
&O\left(T(N+M)^3+TN^3(N+M)\log N\right)\\
&=O\left(T(N^6+N^5\log N)\right)\\
&=O(TN^6),
\end{aligned}
\]
where the simplification follows from $M=O(N^2)$ for simple graphs.

For Dirac space complexity, storing the raw vertex and edge signatures across all $T$ sampled time points requires $O(T(N+M)^2)$ space. The incident-edge signatures and BFS aggregation reuse the stored signature values without duplicating them, requiring at most $O(N^2)$ additional space for references.For space complexity, storing the raw vertex and edge signatures across all $T$ sampled time points requires $O(T(N+M)^2)$ space. The incident-edge signatures and BFS aggregation reuse the stored signature values without duplicating them, requiring at most $O(N^2)$ additional space. Since $M=O(N^2)$ for simple graphs, the overall worst-case space complexity simplifies to
\[
O(T(N+M)^2)=O(TN^4).
\]



As for the correlation matrix, computing the pseudoinverse of the $N\times N$ Laplacian matrix requires $O(N^3)$ time and $O(N^2)$ space. Computing all-pairs shortest-path distances requires $O(N^3)$ time, while sorting the correlation matrix rows according to these distances requires $O(N^2\log N)$ time. Since these operations are bounded by the pseudoinverse and shortest-path computations, constructing the raw correlation signatures requires $O(N^3)$ time.

The resulting signatures are then aggregated according to BFS layers. Performing BFS from every vertex requires $O(N(N+M))=O(N^3)$ time. For each vertex, at most $N$ raw signatures are sorted across its layers, requiring $O(N\log N)$ comparisons. Since each signature contains $O(N)$ values, each comparison may take $O(N)$ time, giving $O(N^2\log N)$ time per vertex. Across all $N$ vertices, the overall worst-case time complexity, including BFS aggregation, is
BFS aggregation, is
\[
O(N^3\log N).
\]
For the correlation space complexity, storing the correlation matrix, shortest-path distances, and intermediate matrices requires $O(N^2)$ space. BFS aggregation organizes the existing raw signatures into distance layers without duplicating their values, requiring at most $O(N^2)$ additional space. Therefore, the overall worst-case space complexity is
\[
O(N^2).
\]

Each distance graph $\mathcal{G}_i$ preserves the original $N$ vertices, with edges connecting pairs of vertices at distance exactly $i$, having at most $\binom{N}{2}=O(N^2)$ edges. Constructing $\mathcal{G}_i$ requires $O(N^2)$ additional time and space. Thus, the overall time and space complexities of constructing a single distance graph are $O(N^3)$ and $O(N^2)$, respectively.

For line graph constructions, each edge of $\mathcal{G}$ becomes a vertex in $L(\mathcal{G})$, giving $|V(L(\mathcal{G}))|=M=O(N^2)$. Two vertices in $L(\mathcal{G})$ are adjacent if their corresponding edges in $\mathcal{G}$ share an endpoint. Thus, each vertex $v\in V$ contributes $\binom{\deg(v)}{2}$ edges to $L(\mathcal{G})$. Since $\mathcal{G}$ is a simple graph, $\deg(v)\leq N-1$ for every vertex. Therefore,
\[
|E(L(\mathcal{G}))|=\sum_{v\in V}\binom{\deg(v)}{2}\leq \sum_{v\in V}\binom{N-1}{2}=N\binom{N-1}{2}=O(N^3).
\] 
More generally, for any fixed $i\geq 0$, the iterated line graph $L^i(\mathcal{G})$ has at most $O(N^{i+1})$ vertices and $O(N^{i+2})$ edges.

We analyze the spectral complexity across graph transformations by considering the successive line graphs $L^i(\mathcal{G})$ and their corresponding distance graphs.
Let $\ell$ denote the maximum line-graph order considered by the algorithm. At each order $0\leq i\leq\ell$, the algorithm applies the spectral operators to $L^i(\mathcal{G})$ and its distance graphs $(L^i(\mathcal{G}))_j$, where $1\leq j\leq D_i$ and $D_i$ is the diameter of $L^i(\mathcal{G})$. We use $j=1$ to represent $L^i(G)$ itself, so $i=0$ and $j=1$ corresponds to the original graph $\mathcal{G}$. Let $N_i$ denote the number of vertices in $L^i(\mathcal{G})$ and $T_{i,j}$ the maximum number of sampled time points across the time-dependent operators at that stage.
Using the previously established bounds for the Laplacian heat kernel, Dirac operator, and correlation matrix, the total time complexity of the spectral evaluations across all transformations is
\[
O\left(\sum_{i=0}^{\ell}\sum_{j=1}^{D_i}
\left[T_{i,j}N_i^3\log N_i+T_{i,j}N_i^6+N_i^3\log N_i\right]\right).
\]
Since the Dirac operator has the highest asymptotic time complexity among the three operators, this simplifies to
\[
O\left(\sum_{i=0}^{\ell}\sum_{j=1}^{D_i}T_{i,j}N_i^6\right).
\]
Using $D_i=O(N_i)$ and $N_i=O(N^{i+1})$, the worst-case time complexity for fixed $\ell$ can be further bounded as
\[
\begin{aligned}
&=O\left(T_{\max}\sum_{i=0}^{\ell}D_iN_i^6\right)=O\left(T_{\max}\sum_{i=0}^{\ell}N_i^7\right)=O\left(T_{\max}\sum_{i=0}^{\ell}N^{7i+7}\right)=O(T_{\max}N^{7\ell+7})
\end{aligned}
\]
where $T_{\max}$ denotes the maximum number of sampled time points across all stages.

Similarly, the Dirac operator has the largest space requirement. Since the operators and graph transformations are processed sequentially, the peak space complexity is determined by the largest spectral computation. Using the previously established Dirac bound, each stage requires at most $O(T_{i,j}N_i^4)$ space. Since $N_i=O(N^{i+1})$, the overall worst-case space complexity for fixed $\ell$ is
\[
O(T_{\max}N^{4\ell+4}).
\]

We analyze the probing techniques applied to the original graph $\mathcal{G}$ following spectral refinement. Let $C\subseteq V$ denote the selected unresolved vertex class, with $|C|=N$ in the worst case. For each probed vertex $u\in C$, a probing technique constructs a set of temporarily modified versions of graphs $\mathcal{G}$, and computes their vertex signatures using the correlation matrix to form the probing profile $P(u)$. Since each modified graph has at most $N$ vertices, computing its signatures requires $O(N^3\log N)=O(N^4)$ time and $O(N^2)$ working space. 

Folding considers $M$ disjoint groups of size $p$, with the first group containing the probed vertex $u$. Since the order of the additional groups and the vertices within each group does not matter, the number of possible modifications per vertex is
\[
F(N,p,M)=\binom{N-1}{p-1}\frac{(N-p)!}{(N-Mp)!(p!)^{M-1}(M-1)!}.
\]
Expanding the binomial coefficient gives
\[
F(N,p,M)=\frac{(N-1)!}{(p-1)!(N-Mp)!(p!)^{M-1}(M-1)!}.
\]
For fixed $p$ and $M$, this gives $O(N^{Mp-1})$ modifications per vertex. Each modified graph has $N-M(p-1)$ vertices, and computing its correlation-based signatures requires at most $O(N^4)$ time. Considering all $N$ vertices, the overall worst-case time complexity is
\[
O(N\cdot N^{Mp-1}\cdot N^4)=O(N^{Mp+4}).
\]

For space complexity, each modification contributes one signature containing at most \(O(N^2)\) values to the probing profile of the selected vertex. Since each vertex has $O(N^{Mp-1})$ possible modifications, storing the profiles for all $N$ vertices requires
\[
O(N\cdot N^{Mp-1}\cdot N^2)=O(N^{Mp+2})
\]
space.

Neighborhood deletion instead selects a group of $p$ vertices including the probed vertex $u$, giving at most $\binom{N-1}{p-1}$ possible modifications. Restrictions on the distances between selected vertices may reduce this number. For fixed $p$, this corresponds to $O(N^{p-1})$ modifications per vertex, or $O(N^p)$ modifications across all $N$ vertices. Since each modified graph has at most $N$ vertices and computing its correlation-based signatures requires $O(N^4)$ time, the overall worst-case time complexity is
\[
O(N^p\cdot N^4)=O(N^{p+4}).
\]

For space complexity, each probing profile stores at most $\binom{N-1}{p-1}=O(N^{p-1})$ signatures, each containing at most $O(N^2)$ entries. Considering all $N$ vertices, the space required to store the probing profiles is
\[
O(N\cdot N^{p-1}\cdot N^2)=O(N^{p+2}).
\]

Node deletion requires only one modification per probed vertex, obtained by removing that vertex and its incident edges. Computing the signatures of the remaining $N-1$ vertices requires $O(N^3)$ time. Sorting these signatures does not increase the asymptotic bound. Since the procedure is repeated for all $N$ vertices, the overall worst-case time complexity is
\[
O(N\cdot N^4)=O(N^5).
\]

For space complexity, each probing profile stores the BFS-aggregated signatures of the $N-1$ remaining vertices. Each signature contains at most $O(N)$ references to raw correlation signatures, whose underlying values are shared rather than duplicated. Therefore, storing one probing profile requires $O(N^2)$ space, and retaining the profiles for all $N$ probed vertices requires $O(N^3)$ space.

\section{Results}
To evaluate the effectiveness of the proposed hierarchy in recovering automorphism orbits, we compare the partition induced by the computed signatures with the true automorphism-orbit partition of each graph, obtained using nauty and Traces~\cite{mckay2014practical}. We use two evaluation metrics: normalized mutual information (NMI) and an orbit-compatibility score $P$. NMI measures the overall similarity between the computed and true orbit partitions, with a value of $1$ indicating an exact match. However, NMI does not directly capture the usefulness of the computed classes for graph isomorphism testing, which is an intended future application of our signature library.
To complement NMI, we introduce an orbit-compatibility score $P$, which evaluates how reliably vertices grouped into the same signature class belong to the same true automorphism orbit. Specifically, for a given signature class, we consider the probability that two vertices independently and uniformly selected from that class belong to the same orbit.


We evaluated the proposed hierarchy on benchmark graphs drawn from several structurally distinct families, including grids, Hadamard and projective-plane graphs, Steiner triple-system graphs, and the MZ, USR, Latin-SW, CHH, and F-LEX families. 
The simplest stage of the hierarchy already suffices for many instances, without recourse to graph transformations or probing. For example, the $\mathcal{G}$-Laplacian signature recovers the 19 orbit classes of \texttt{kef10}, which has 200 vertices, and the 325 classes of \texttt{grid-2-50}, which has 2,500 vertices. The appendix documents analogous exact recovery for numerous members of the grid, CHH, TNN, and other benchmark families. These examples demonstrate that the more expensive stages are not invariably necessary. At the same time, success on a graph with only one orbit, such as a vertex-transitive graph, should not be confused with distinguishing power: the more demanding tests are those in which several true orbits are initially merged by the signatures.

Table~\ref{tab:representative-orbit-results} collects a selection of such instances. The initial number of classes is that obtained from the $\mathcal{G}$-Laplacian signature, and the listed method is the first stage in the tested sequence at which the exact orbit partition is recovered. We display the number of recovered orbits and the corresponding $k$-WL result, while the complete progression of intermediate classes, NMI and $P$ values remains in the appendix.

\begin{table}[t]
\centering
\small
\setlength{\tabcolsep}{3.5pt}
\begin{tabular}{@{}lrrlrc@{}}
\toprule
\textbf{Graph} & $\mathbf{N}$ & \textbf{Initial classes} & \textbf{First successful stage} & \textbf{Orbits} & $\mathbf{k}$\textbf{-WL} \\
\midrule
\texttt{usr(1)\_29-1}   & 29  & 1   & $\mathcal{G}$-Dirac                     & 13  & 4 \\
\texttt{sts-13}        & 26  & 1   & $\mathcal{G}$-Dirac                     & 2   & 4 \\
\texttt{sts-15}        & 35  & 1   & $\mathcal{G}^{(2)}$-Dirac               & 2   & 4 \\
\texttt{latin-sw-10-2} & 100 & 1   & $\mathcal{G}^{(2)}$-Dirac               & 100 & -- \\
\texttt{mz-4}          & 80  & 16  & $\mathcal{G}^{(3)}$-Laplacian           & 20  & 3 \\
\texttt{had-52}        & 208 & 1   & $\mathcal{G}$-folding                   & 2   & $>3$ \\
\texttt{pp-9-2}        & 182 & 1   & $\mathcal{G}$-neighborhood deletion     & 4   & $>3$ \\
\texttt{had-sw-32-1}   & 128 & 1   & $\mathcal{G}$-neighborhood deletion     & 42  & $>3$ \\
\texttt{f-lex-reg-10-1} & 340 & 141 & $\mathcal{G}$-neighborhood deletion    & 175 & -- \\
\bottomrule
\end{tabular}
\caption{Representative benchmark instances requiring different stages of the signature hierarchy. ``Initial classes'' refers to the $\mathcal{G}$-Laplacian partition; ``orbits'' gives the number of classes in the recovered orbit partition. All listed successful stages attain ${\rm NMI}=P=1$. The $k$-WL column gives the smallest tested dimension recovering the orbit partition; $>3$ denotes failure for $k\leq 3$, and -- denotes an unreported value. The appendix gives the full results and probe parameters.}
\label{tab:representative-orbit-results}
\end{table}

The distinctions in Table~\ref{tab:representative-orbit-results} are informative about what each additional stage contributes. For \texttt{usr(1)\_29-1} and \texttt{sts-13}, the Laplacian and correlation signatures each yield a single class, whereas the Dirac stage recovers, respectively, 13 and two orbits. For \texttt{sts-15}, neither the original-graph Dirac signature nor the Laplacian and correlation signatures of the distance-two graph distinguish its two orbits; the distance-two Dirac signature does. In \texttt{latin-sw-10-2}, the original-graph Dirac signature yields 99 classes instead of the 100 singleton orbits, and the distance-two Dirac stage resolves the remaining pair. A different benefit of distance transformations appears in \texttt{mz-4}: the first two distance levels leave 16 classes, while the distance-three Laplacian signature recovers all 20 orbits. Thus, increasing the information retained by the signatures can yield successive, and sometimes very fine, refinements of the partition.

The most demanding examples in the table require a change of strategy. In \texttt{had-52}, the combined spectral signatures leave all 208 vertices in one class, whereas folding pairs of vertices separates its two orbits. In \texttt{pp-9-2}, the spectral constructions on the original graph, its distance graphs, and the tested line-graph transformations likewise leave a single class; neighborhood deletion with $p=3$ and distance restriction $d=2$ recovers all four orbits. Neighborhood deletion also resolves the 42 orbits of \texttt{had-sw-32-1} and refines the 141 initial classes of \texttt{f-lex-reg-10-1} into its 175 true orbits. These examples illustrate why probing is not merely an optional embellishment of the spectral hierarchy: in the tested instances, controlled modifications reveal distinctions that remain hidden in the unmodified graphs.

The comparison with $k$-WL addresses the same vertex-orbit recovery task. The reported $k$-WL values show that \texttt{usr(1)\_29-1}, \texttt{sts-13}, and \texttt{sts-15} require dimension four among those tested, while \texttt{had-52}, \texttt{pp-9-2}, and \texttt{had-sw-32-1} are not resolved at any of dimensions one through three. Nevertheless, their exact orbit partitions are recovered by the spectral or probing stages indicated in Table~\ref{tab:representative-orbit-results}. This establishes the distinguishing power of the proposed signatures on particular examples that resist low-dimensional WL refinement. It does not imply a universal superiority over $k$-WL, nor does successful recovery alone establish an advantage in running time.

Finally, the observed effectiveness must be read alongside the numerical and computational limitations of the method. The detailed tables include numerical resolutions, so the reported partitions depend on the stated comparison settings. Moreover, Dirac computations, graph transformations, and exhaustive probing may be substantially more expensive than the initial Laplacian stage, as discussed in the complexity analysis. The experiments therefore support a hierarchy that succeeds on the tested benchmark instances, including some for which low-dimensional $k$-WL fails, without establishing completeness on arbitrary graphs or a general polynomial-time orbit-identification algorithm.

\section{Conclusion}
We return to the question with which we began. On a network, the automorphism group is the group of transformations that leave the interaction structure invariant, and its orbits are the classes of vertices that no symmetry can distinguish. Recovering those orbits is therefore the precise form of a physical question: which parts of a network are genuinely equivalent, and which merely look alike to the local probes available? The hierarchy we have developed answers this question incrementally, separating what can be resolved by spectral data from what requires controlled intervention in the graph. Its success on instances that resist WL refinement is evidence that the symmetry-based description of a physical system, which in statistical physics organizes phases and collective behavior, can be made operational for discrete, complex networks, not as a single diagnostic, but as a graded sequence of tests whose information content can be read from the partition it produces. The limitations are equally part of the result: the method is empirically successful on the tested families, its later stages are computationally non-trivial, and a general polynomial-time procedure for orbit identification remains out of reach. These are the boundaries within which the symmetry-based description of networks is, at present, a practical tool.
\section{Acknowledgements}
Mohamad Al Sayed received support as a research assistant from the Complexity and Network Science Cluster, Rana Dbouk also received support through the Complexity and Network Science Cluster and the University Research Board. Joe Germany is grateful for helpful discussions with Prof. Siamak Taati.

\printbibliography

\appendix
\section{Proofs}
\label{app:proofs}

In the below, let $\mathcal{G}$ be a finite simple undirected graph on $n$ vertices, with
Laplacian $L = D - A$. We write the heat kernel and the correlation matrices as
\begin{equation}
    K(t) = e^{-tL}, \qquad C_r=(L+rI)^{-1},
\end{equation}
with the softening parameter $r > 0$.

\subsection{Counterexamples}
\label{app:counterexamples}

We first prove a counterexample to our first algorithm, which involving only the values of the diagonal.
\begin{theorem}
\label{thm:frucht-diagonal}
Let $\mathcal{G}$ be the Frucht graph with the vertex labeling returned by
\texttt{nx.frucht\_graph()}. Vertices $1$ and $10$ belong to different
orbits, but
\begin{equation}
    K_{1,1}(t)=K_{10,10}(t),\qquad t\geq0.
\end{equation}
Thus, the diagonal heat-kernel signatures do not distinguish their orbits.
\end{theorem}
\begin{proof}
The graph has $12$ vertices. Computing $L^0,\ldots,L^{11}$ gives
\begin{equation}
\begin{aligned}
    \bigl((L^m)_{1,1}\bigr)_{m=0}^{11}
    &=\bigl((L^m)_{10,10}\bigr)_{m=0}^{11}\\
    &=(1,3,12,52,234,1078,5054,24034,\\
    &\hspace{2em}115660,562230,2756552,13614304).
\end{aligned}
\end{equation}
By Cayley--Hamilton, for every integer $s\geq0$, there are coefficients
$c_{s,0},\ldots,c_{s,11}$ such that
\begin{equation}
    L^s=\sum_{m=0}^{11}c_{s,m}L^m.
\end{equation}
Consequently,
\begin{equation}
    (L^s)_{1,1}-(L^s)_{10,10}
    =\sum_{m=0}^{11}c_{s,m}
      \bigl((L^m)_{1,1}-(L^m)_{10,10}\bigr)=0.
\end{equation}
Using the exponential representation of the heat kernel, we obtain
\begin{equation}
    K_{1,1}(t)-K_{10,10}(t)
    =\sum_{s=0}^{\infty}\frac{(-t)^s}{s!}
      \bigl((L^s)_{1,1}-(L^s)_{10,10}\bigr)=0.
\end{equation}
To distinguish the orbits, let $T$ be the set of vertices that belong to
no triangle. For this labeling, we have
\begin{equation}\begin{aligned}
    T&=\{2,8,11\},\\
    \{d(1,j):j\in T\}&=\{1,2,2\},\\
    \{d(10,j):j\in T\}&=\{1,2,3\}.
\end{aligned}\end{equation}
An automorphism preserves triangles and distances, and therefore maps
$T$ onto itself. If it mapped $1$ to $10$, the last two collections
would be equal. Hence $1$ and $10$ belong to different orbits.
\end{proof}

This next proof proves a counterexample to our algorithm that treats signatures as the full sorted rows of the heat kernel, without BFS.
\begin{theorem}
\label{thm_sts13_sorted_rows}
Let $\mathcal{G}$ be the \texttt{sts-13} graph on $26$ vertices,
with labels $0,\ldots,25$, obtained from the nauty dataset \cite{nautytracesgraphs}. Vertices $0$ and $3$ belong to different orbits, but
\begin{equation}
\label{sts13_sorted_heat_rows}
    \operatorname{sort}\bigl(K_{0j}(t):j\in V\bigr)
    =\operatorname{sort}\bigl(K_{3j}(t):j\in V\bigr),
    \qquad t\geq0.
\end{equation}
\end{theorem}
\begin{proof}
Let $\pi$ be the permutation given by
\begin{equation}
\label{sts13_permutation}
\begin{aligned}
    \bigl(\pi(0),\ldots,\pi(25)\bigr)
    ={}&(3,0,1,2,4,5,7,9,6,8,12,10,13,\\
        &\phantom{(}11,14,15,18,21,22,16,23,17,19,20,25,24).
\end{aligned}
\end{equation}
Computing the first $26$ powers of $L$ gives
\begin{equation}
\label{sts13_finite_power_identity}
    (L^m)_{0j}=(L^m)_{3,\pi(j)},
    \qquad j\in V,\quad m=0,\ldots,25.
\end{equation}
The same permutation $\pi$ is used for every power. By Cayley--Hamilton,
for every integer $s\geq0$, we can write
\begin{equation}
    L^s=\sum_{m=0}^{25}c_{s,m}L^m.
\end{equation}
Thus, for every $j\in V$,
\begin{equation}
\label{sts13_all_power_identity}
    (L^s)_{0j}-(L^s)_{3,\pi(j)}
    =\sum_{m=0}^{25}c_{s,m}
      \bigl((L^m)_{0j}-(L^m)_{3,\pi(j)}\bigr)=0.
\end{equation}
It follows that
\begin{equation}
\label{sts13_heat_identity}
    K_{0j}(t)-K_{3,\pi(j)}(t) =\sum_{s=0}^{\infty}\frac{(-t)^s}{s!} \bigl((L^s)_{0j}-(L^s)_{3,\pi(j)}\bigr) = 0.
\end{equation}
Since $\pi$ is a permutation, sorting the two rows gives
\eqref{sts13_sorted_heat_rows}.

Now let $q(i)$ be the number of four-vertex cliques containing $i$.
Each such clique consists of $i$ and three pairwise adjacent neighbors.
Enumerating these triples gives
\begin{equation}
\label{sts13_clique_counts}
    q(0)=73,\qquad q(3)=71.
\end{equation}
An automorphism mapping $0$ to $3$ would map the cliques containing
$0$ bijectively onto those containing $3$. Since the counts differ,
the two vertices belong to different orbits.
\end{proof}

The same \texttt{sts-13} graph is still a counterexample to the algorithm that incorporates both inner and outer BFS signatures.
\begin{theorem}
\label{thm_sts13_bfs}
Let $\mathcal{G}$ be the same labeled \texttt{sts-13} graph. Its inner
and outer BFS heat-kernel signatures satisfy
\begin{equation}
\begin{aligned}
    \operatorname{sig}_{\mathcal{G},\mathrm{inter}}(v)(t)
      &=\operatorname{sig}_{\mathcal{G},\mathrm{inter}}(0)(t),
      &&v\in V,\quad t\geq0,\\
    \operatorname{sig}_{\mathcal{G}}(v)
      &=\operatorname{sig}_{\mathcal{G}}(0),
      &&v\in V,
\end{aligned}
\end{equation}
for any common choice of sampling times and outer BFS depth.
In particular, both signatures assign equal values to vertices $0$
and $3$, although these vertices belong to different orbits.
\end{theorem}

\begin{proof}
For every $v\in V$, the exact computation gives a permutation
$\pi_v:V\to V$ such that
\begin{equation}
\label{sts13_all_root_power_matching}
\begin{aligned}
    d(0,j)&=d(v,\pi_v(j)),\\
    (L^m)_{0j}&=(L^m)_{v,\pi_v(j)},
    \qquad m=0,\ldots,25,
\end{aligned}
\end{equation}
for every $j\in V$. These permutations are obtained by matching the
vectors $\bigl((L^m)_{0j}\bigr)_{m=0}^{25}$ with the corresponding
vectors at $v$, within equal distance layers. The accompanying
script checks both equalities directly. Each $\pi_v$ is fixed for
all $26$ powers. For $v=3$, we may take the permutation in
\eqref{sts13_permutation}.

By Cayley--Hamilton, for every integer $s\geq0$, there are
coefficients $c_{s,0},\ldots,c_{s,25}$ such that
\begin{equation}
    L^s=\sum_{m=0}^{25}c_{s,m}L^m.
\end{equation}
Consequently,
\begin{equation}
    (L^s)_{0j}-(L^s)_{v,\pi_v(j)}
    =\sum_{m=0}^{25}c_{s,m}
      \bigl((L^m)_{0j}-(L^m)_{v,\pi_v(j)}\bigr)=0.
\end{equation}
Using the exponential representation of the heat kernel gives
\begin{equation}
\label{eq:sts13-bfs-kernel-matching}
\begin{aligned}
    K_{0j}(t)-K_{v,\pi_v(j)}(t)
    &=\sum_{s=0}^{\infty}\frac{(-t)^s}{s!}
      \bigl((L^s)_{0j}-(L^s)_{v,\pi_v(j)}\bigr)\\
    &=0,\qquad t\geq0.
\end{aligned}
\end{equation}

First, consider the inner BFS signature. Since
$d(v,\pi_v(0))=d(0,0)=0$, we have $\pi_v(0)=v$, and hence
\begin{equation}
    K_{00}(t)=K_{vv}(t).
\end{equation}
Moreover, $\pi_v$ maps the distance-$d$ layer of $0$ bijectively
onto the distance-$d$ layer of $v$. Thus, for every $d\geq1$,
\begin{equation}
\begin{aligned}
    \{K_{0j}(t):d(0,j)=d\}
    &=\{K_{v,\pi_v(j)}(t):d(0,j)=d\}\\
    &=\{K_{v\ell}(t):d(v,\ell)=d\}.
\end{aligned}
\end{equation}
The diagonal entries agree, and the entries in each corresponding
layer agree with multiplicity. Sorting within the layers and
concatenating them therefore gives
\begin{equation}
    \operatorname{sig}_{\mathcal{G},\mathrm{inter}}(0)(t)
      =\operatorname{sig}_{\mathcal{G},\mathrm{inter}}(v)(t),
      \qquad v\in V,\quad t\geq0.
\end{equation}
For any common sampling times $t_0,\ldots,t_T$, it follows that
\begin{equation}
\begin{aligned}
    \operatorname{sig}_{\mathcal{G},\mathrm{inter}}(0)
    &=\bigl(\operatorname{sig}_{\mathcal{G},\mathrm{inter}}(0)(t_a)
       \bigr)_{a=0}^{T}\\
    &=\bigl(\operatorname{sig}_{\mathcal{G},\mathrm{inter}}(v)(t_a)
       \bigr)_{a=0}^{T}
     =\operatorname{sig}_{\mathcal{G},\mathrm{inter}}(v).
\end{aligned}
\end{equation}
In particular, the inner BFS signatures of $0$ and $3$ agree.

Now consider the outer BFS signature. The preceding equality holds
at every vertex, so, for every $j\in V$,
\begin{equation}
    \operatorname{sig}_{\mathcal{G},\mathrm{inter}}(j)
      =\operatorname{sig}_{\mathcal{G},\mathrm{inter}}(\pi_v(j)).
\end{equation}
Using the same distance-layer bijection $\pi_v$, we obtain
\begin{equation}
\begin{aligned}
    &\{\operatorname{sig}_{\mathcal{G},\mathrm{inter}}(j):d(0,j)=d\}\\
    &\qquad=\{\operatorname{sig}_{\mathcal{G},\mathrm{inter}}(\pi_v(j)):
                     d(0,j)=d\}\\
    &\qquad=\{\operatorname{sig}_{\mathcal{G},\mathrm{inter}}(\ell):
                     d(v,\ell)=d\}.
\end{aligned}
\end{equation}
Thus each corresponding outer layer contains the same number of
copies of the common intermediary signature. For any depth $b$,
\begin{equation}
\begin{aligned}
    \operatorname{sig}_{\mathcal{G}}(0)
    &=\bigl(\operatorname{sort}(
        \operatorname{sig}_{\mathcal{G},\mathrm{inter}}(j):d(0,j)=d)
       \bigr)_{d=0}^{b}\\
    &=\bigl(\operatorname{sort}(
        \operatorname{sig}_{\mathcal{G},\mathrm{inter}}(\ell):d(v,\ell)=d)
       \bigr)_{d=0}^{b}
     =\operatorname{sig}_{\mathcal{G}}(v).
\end{aligned}
\end{equation}
This includes the root layer and any empty layers beyond the
maximum distance. Hence the outer BFS signatures of $0$ and $3$
also agree. However, \eqref{sts13_clique_counts} gives $q(0)=73$
and $q(3)=71$, so the two vertices belong to different orbits.
Therefore both the inner and outer BFS signatures fail to recover
the vertex-orbit partition of $\mathcal{G}$.
\end{proof}

The same counterexamples hold for the corresponding correlation
signatures. Indeed, each entry equality above holds under a fixed
matching for every $t\geq0$, so
\begin{equation}
    (C_r)_{ij}-(C_r)_{k\ell}
    =\int_0^\infty e^{-rt}
      \bigl(K_{ij}(t)-K_{k\ell}(t)\bigr)\,dt=0,
    \qquad r>0.
\end{equation}
The same matchings therefore apply within every BFS layer.

They also apply to the pseudoinverse $L^\dagger$ used by the
implementation. For any finite graph, let $\lambda_1,\ldots,\lambda_q$
be its distinct nonzero Laplacian eigenvalues. Polynomial
interpolation gives a polynomial $p$ of degree at most $q\leq n-1$
with $p(0)=0$ and $p(\lambda_a)=1/\lambda_a$. By the spectral theorem,
\begin{equation}
    L^\dagger=p(L)=\sum_{m=0}^{n-1}a_m L^m.
\end{equation}
Every finite-power equality verified above therefore gives the
corresponding equality of pseudoinverse entries. This uses no
additional property of the counterexample graphs. It does not
assert that the single matrix $L^\dagger$ has the same distinguishing
power as the family $(L+rI)^{-1}$ for all $r>0$.

Finally, the other graphs (sts-15, pp-9-2, had-52) that guide the storyline above can also be explicitly proved to be counterexamples to each iteration of the algorithm based on the same proof machinery.

\subsection{One direction of invariance}
\label{app:oneDirection}

In this section, we show that vertices in the same orbit of
$\mathcal{G}$ receive equal signatures.

We first show that vertices in the same orbit of $\mathcal{G}$
remain in the same orbit of each distance graph $\mathcal{G}^{(i)}$.

\begin{proposition}
\label{prop:distance-orbits}
\label{orbits_under_perturbation}
Let $\mathcal{G}=(V,E)$ be a graph on $n$ vertices, and let $u,v\in V$.
If $u$ and $v$ are in the same orbit in $\mathcal{G}$, then they are in the
same orbit in $\mathcal{G}^{(i)}$ for all $i\in\{1,\ldots,n\}$.
\end{proposition}

\begin{proof}
If $u$ and $v$ belong to the same vertex orbit of the graph $\mathcal{G}$,
then there exists an automorphism $f$ such that $f(u)=v$. We claim
that $f$ is also an automorphism of $\mathcal{G}^{(i)}$ for every
$i\in\{1,\ldots,n\}$.

Let $x,y\in V$ be connected by a path. Since $f$ preserves
adjacency, it maps a shortest path from $x$ to $y$ to a path of
the same length from $f(x)$ to $f(y)$. Therefore,
\begin{equation}
    d_{\mathcal{G}}(f(x),f(y))\leq d_{\mathcal{G}}(x,y).
\end{equation}
Applying the same argument to $f^{-1}$ gives
\begin{equation}
    d_{\mathcal{G}}(x,y)\leq d_{\mathcal{G}}(f(x),f(y)).
\end{equation}
Thus,
\begin{equation}
\label{automorphism_distance_preservation}
    d_{\mathcal{G}}(f(x),f(y))=d_{\mathcal{G}}(x,y).
\end{equation}
If $x$ and $y$ are disconnected, so are $f(x)$ and $f(y)$;
otherwise, $f^{-1}$ would map a path between their images to a
path between $x$ and $y$.

By the definition of $\mathcal{G}^{(i)}$, two vertices are adjacent in
$\mathcal{G}^{(i)}$ if and only if their distance in $\mathcal{G}$ is $i$. Hence,
\begin{equation}
\begin{aligned}
    \{x,y\}\in E(\mathcal{G}^{(i)})
    &\iff d_{\mathcal{G}}(x,y)=i\\
    &\iff d_{\mathcal{G}}(f(x),f(y))=i\\
    &\iff\{f(x),f(y)\}\in E(\mathcal{G}^{(i)}).
\end{aligned}
\end{equation}
Since $f$ is a bijection of $V$, it is an automorphism of
$\mathcal{G}^{(i)}$. Moreover, $f(u)=v$, so $u$ and $v$ belong to the same
orbit in $\mathcal{G}^{(i)}$, for every $i\in\{1,\ldots,n\}$.
\end{proof}

An individual distance graph can therefore merge original orbits,
but cannot split them. For example, the path $0$--$1$--$2$--$3$ has
orbits $\{0,3\}$ and $\{1,2\}$, whereas its distance-$2$ graph consists
of the two edges $\{0,2\}$ and $\{1,3\}$ and has a single vertex orbit.
The permutations preserving the whole family, including
$\mathcal{G}^{(1)}$, are exactly the automorphisms of $\mathcal{G}$.

\begin{theorem}
\label{thm_signature_invariance}
Let $\mathcal{G}=(V,E)$ and $u,v\in V$. If $u$ and $v$ belong to
the same orbit in $\mathcal{G}$, then
\begin{equation}
    \operatorname{sig}_{\mathcal{G}}(u)
      =\operatorname{sig}_{\mathcal{G}}(v).
\end{equation}
This holds for the heat-kernel, correlation, and Dirac signatures,
including their combinations on distance and iterated line graphs,
with common settings, sampling times, and comparison parameters.
The matrix entries are exact before rounding or truncation, and
any relative-norm threshold is positive. The Dirac signatures are
independent of the chosen edge orientations.
\end{theorem}

\begin{proof}
If $u$ and $v$ belong to the same vertex orbit of $\mathcal{G}$,
then there exists an automorphism $f$ such that $f(u)=v$. We claim
that $\operatorname{sig}_{\mathcal{G}}(x)
=\operatorname{sig}_{\mathcal{G}}(f(x))$ for every $x\in V$.

Since $f$ preserves adjacency and degrees,
$L_{f(x),f(y)}=L_{xy}$. Induction on $m$ gives
$(L^m)_{f(x),f(y)}=(L^m)_{xy}$, and hence
\begin{equation}
\label{automorphism_heat_entries}
    K_{f(x),f(y)}(t)
      =\sum_{m=0}^{\infty}\frac{(-t)^m}{m!}(L^m)_{f(x),f(y)}
      =K_{xy}(t).
\end{equation}
The same entry equality holds for the correlation $L^\dagger$
used in the code, since $L^\dagger$ is a polynomial in $L$, and
for $C_r=(L+rI)^{-1}$, $r>0$, by uniqueness of the inverse.
Moreover, $f$ maps each distance layer around $x$ bijectively onto
the corresponding layer around $f(x)$. Thus the diagonal, sorted
full-row, and inner BFS constructions give
\begin{equation}
    \operatorname{sig}_{\mathcal{G},\mathrm{inter}}(x)(t)
      =\operatorname{sig}_{\mathcal{G},\mathrm{inter}}(f(x))(t).
\end{equation}
This remains true after common scalar rounding or truncation and
concatenation in time order.

For the Dirac construction, extend $f$ to edges by
$f(\{a,b\})=\{f(a),f(b)\}$. Set $\epsilon_x=1$ for $x\in V$
and $\epsilon_e=1$ or $-1$ according as $f$ preserves or reverses
the chosen orientation of $e$. The incidence matrix then satisfies
$B_{f(x),f(e)}=\epsilon_e B_{xe}$. Therefore
\begin{equation}
\label{eq:dirac-power-signs}
    (\mathcal D^m)_{f(x),f(y)}
      =\epsilon_x\epsilon_y(\mathcal D^m)_{xy},
    \qquad x,y\in V\sqcup E,\quad m\geq0.
\end{equation}
Indeed, the equality holds for $m=0,1$, and each intermediate
index $z$ in matrix multiplication contributes $\epsilon_z^2=1$.
For $U(t)=e^{\mathrm{i}\mathcal D t}$, it follows that
$U_{f(x),f(y)}(t)=\epsilon_x\epsilon_y U_{xy}(t)$, so
\begin{equation}
\label{eq:dirac-absolute-invariance}
\begin{aligned}
    |\operatorname{Re}U_{f(x),f(y)}(t)|
      &=|\operatorname{Re}U_{xy}(t)|,\\
    |\operatorname{Im}U_{f(x),f(y)}(t)|
      &=|\operatorname{Im}U_{xy}(t)|.
\end{aligned}
\end{equation}
Consequently, the pairs used in the code satisfy
\begin{equation}
\label{eq:dirac-components}
\begin{aligned}
    p_t(x,y)&=\bigl(
      \operatorname{round}(R|\operatorname{Re}U_{xy}(t)|),
      \operatorname{round}(R|\operatorname{Im}U_{xy}(t)|)\bigr),\\
    p_t(f(x),f(y))&=p_t(x,y).
\end{aligned}
\end{equation}
The map $f$ permutes the vertex and edge blocks separately.
Sorting these pairs within each block and concatenating in time
order therefore gives equal vertex signatures at $x,f(x)$ and
equal edge signatures at $e,f(e)$. Since $x\in e$ if and only if
$f(x)\in f(e)$, appending the sorted temporal signatures of incident
edges also preserves equality. These Dirac blocks use no inner BFS.
Using $e^{-\mathrm{i}\mathcal D t}$ gives the same pairs, since
it is the complex conjugate of $U(t)$.

For the Dirac correlation, the matrix
\begin{equation}
    \mathcal D_\mu=
    \begin{pmatrix}\mu I&B\\B^{\mathsf T}&-\mu I\end{pmatrix},
    \qquad \mu\in\mathbb R\setminus\{0\},
\end{equation}
is invertible because its square has positive definite diagonal
blocks $\mu^2I+BB^{\mathsf T}$ and $\mu^2I+B^{\mathsf T}B$.
It satisfies the same sign rule as $\mathcal D$, and uniqueness
of the inverse gives
\begin{equation}
    (\mathcal D_\mu^{-1})_{f(x),f(y)}
      =\epsilon_x\epsilon_y(\mathcal D_\mu^{-1})_{xy}.
\end{equation}
Thus the same absolute-value and sorting argument applies.
Changing edge orientations multiplies each matrix entry by the
corresponding row and column signs. The preceding calculation,
without permuting indices, shows that the absolute real and imaginary
parts, and hence both Dirac constructions, remain unchanged.

By Proposition~\ref{prop:distance-orbits}, $f$ is an automorphism
of every distance graph. It also induces an automorphism of the
line graph, since
\begin{equation}
    e\cap e'\ne\varnothing
      \quad\iff\quad f(e)\cap f(e')\ne\varnothing.
\end{equation}
Repeating these arguments gives an automorphism $f_H$ of every
transformed graph $H$.

The membership sets initially satisfy $M_{\mathcal{G}}(x)=\{x\}$.
A distance operation leaves them unchanged, and a line operation gives
\begin{equation}
\label{eq:membership-update}
    M_{\mathcal L(H)}(x)
      =\{\{a,b\}\in E(H):a\in M_H(x)\text{ or }b\in M_H(x)\}.
\end{equation}
We claim that
\begin{equation}
\label{eq:membership-invariance}
    f_H(M_H(x))=M_H(f(x)).
\end{equation}
This holds initially and is preserved by a distance operation.
For a line operation, the induction hypothesis gives
\begin{equation}
\begin{aligned}
    \{a,b\}\in M_{\mathcal L(H)}(x)
    &\iff a\in M_H(x)\text{ or }b\in M_H(x)\\
    &\iff f_H(a)\in M_H(f(x))\text{ or }f_H(b)\in M_H(f(x))\\
    &\iff\{f_H(a),f_H(b)\}\in M_{\mathcal L(H)}(f(x)).
\end{aligned}
\end{equation}
Thus the claim holds after every transformation.

The preceding operator arguments on $H$ give
$\operatorname{sig}_{H,\mathrm{inter}}(y)
=\operatorname{sig}_{H,\mathrm{inter}}(f_H(y))$.
Together with \eqref{eq:membership-invariance}, this yields
\begin{equation}
\begin{aligned}
&\operatorname{sort}\bigl(\operatorname{sig}_{H,\mathrm{inter}}(y):y\in M_H(x)\bigr)
=\operatorname{sort}\bigl(\operatorname{sig}_{H,\mathrm{inter}}(z):z\in M_H(f(x))\bigr).
\end{aligned}
\end{equation}
Concatenating over the prescribed settings and operators therefore gives
\begin{equation}
    \operatorname{sig}_{\mathcal{G},\mathrm{inter}}(x)
      =\operatorname{sig}_{\mathcal{G},\mathrm{inter}}(f(x)).
\end{equation}
Equal arrays receive the same identifier, including in relative-norm
mode, where their comparison distance is zero.

Finally, $f$ preserves distances in the original graph, so for any
common outer BFS depth $b$,
\begin{equation}
\label{eq:implemented-outer-bfs}
\begin{aligned}
    \operatorname{sig}_{\mathcal{G}}(x)
    &=\bigl(\operatorname{sort}(\operatorname{sig}_{\mathcal{G},\mathrm{inter}}(y):d_{\mathcal{G}}(x,y)=d)\bigr)_{d=0}^{b}
    =\operatorname{sig}_{\mathcal{G}}(f(x)).
\end{aligned}
\end{equation}
If outer BFS is disabled, the intermediary signatures already agree.
Taking $x=u$ and using $f(u)=v$ proves the claim.
\end{proof}

The scalar quantization and comparison parameters must be common
to all compared vertices. Floating-point errors in the matrix
computations are not covered by the exact equalities above.

\subsection{Heat kernel and correlation equivalence}
\label{app:heatKernelAndCorrelation}

We prove the following equivalence between the heat kernel for all time $t \geq 0$ and the correlation for $r > 0$:
\begin{theorem}
    For any $i,j,k,\ell \in V$, the following statements are equivalent:
    \begin{enumerate}
        \item[(i)] $K_{ij}(t) = K_{k\ell}(t)$ for all time $t \geq 0$,
        \item[(ii)] $(C_r)_{ij} = (C_r)_{k\ell}$ for every $r > 0$,
        \item[(iii)] $(L^m)_{ij} = (L^m)_{k\ell}$ for $m = 0, \dots, n-1$.
    \end{enumerate}
\end{theorem}
\begin{proof}
We note that, since $L$ is positive semidefinite, $C_r$ is well defined for every
$r>0$; it is the resolvent of $-L$ at $r$.

First, assume (i). By the Laplace transform representation of the resolvent \cite[Lemma 1.3.7]{arora2025eventual}, we can write $C_r$ as
\begin{equation}
    C_r = \int_0^\infty e^{-rt} K(t) \, dt.
\end{equation}
Taking entries gives
\begin{equation}
    (C_r)_{ij} - (C_r)_{k \ell} = \int_0^\infty e^{-rt} (K_{ij} (t) - K_{k \ell}(t)) \, dt,
\end{equation}
from which it is obvious that (i) implies (ii).

Now, assume (ii). For $r > \lambda_\text{max}$, the maximum eigenvalue of $L$, the resolvent can be represented as the Neumann series 
\begin{equation}
    C_r = (L + rI)^{-1} = \sum_{m=0}^\infty \frac{(-1)^m}{r^{m+1}} L^m,
\end{equation}
which convergences absolutely \cite[Proposition 1.2.7(b)]{arora2025eventual}. Thus, by (ii), for $r > \lambda_\text{max}$, we have
\begin{equation}
    0 = (C_r)_{ij} - (C_r)_{k \ell} = \sum_{m=0}^\infty \frac{(-1)^m}{r^{m+1}} \left( (L^m)_{ij} - (L^m)_{k \ell} \right).
\end{equation}
The series vanishes for all sufficiently small values of $1/r$, so every coefficient is zero. In particular, $(L^m)_{ij} = (L^m)_{k \ell}$ for all $m = 0, \dots, n-1$, proving (iii).

Finally, assume (iii). By Cayley-Hamilton, every power $L^m$ is a linear combination of $I, L, \dots, L^{n-1}$. For any $s \geq 0$, we have
\begin{equation}
    (L^s)_{ij} - (L^s)_{k\ell} = \sum_{m=0}^{n-1} c_m \left( (L^m)_{ij} - (L^m)_{k\ell} \right) = 0,
\end{equation}
by the assumption. Thus, by the exponential representation of the heat kernel, we obtain
\begin{equation}
    K_{ij}(t) - K_{k \ell}(t) = \sum_{s=0}^\infty \frac{(-t)^s}{s!} \left( (L^s)_{ij} - (L^s)_{k \ell} \right) = 0, 
\end{equation}
which is (i).
\end{proof}

\section{Additional probing techniques}
\label{app:probing}

\begin{itemize}
\item \textbf{Edge swapping.}
Given two edges $(a,b),(c,d)\in E$ with four distinct endpoints, an edge swap removes these edges and replaces them with either $(a,c),(b,d)$ or $(a,d),(b,c)$. The resulting edge set is
\[
E'=(E\setminus\{(a,b),(c,d)\})\cup
\begin{cases}
\{(a,c),(b,d)\},\\
\{(a,d),(b,c)\}.
\end{cases}
\]
The parameter $M\geq 1$ specifies the number of simultaneous swaps. To probe a vertex $u$, all valid combinations of $M$ swaps are considered, requiring at least one removed edge to be incident to $u$. Swaps must involve distinct removed and added edges, with no edge removed in one swap and added in another. Furthermore, newly added edges must not exist in the original graph, and only modifications preserving connectivity are retained. The signatures $\tau_{G'}(u)$ from all valid modified graphs are collected to construct $P(u)$.

\item \textbf{Single-clique addition.}
Introduce $c\geq 1$ new vertices forming a clique, and connect each of them to $u$. The probing profile $P(u)$ consists of the signature $\tau_{G'}(u)$.

\item \textbf{Edge addition.}
Add $M\geq 1$ non-existing edges simultaneously, requiring at least one to be incident to $u$. All possible combinations of $M$ edges satisfying this condition are considered, and the signatures $\tau_{G'}(u)$ of the resulting graphs are collected to construct $P(u)$.

\item \textbf{Edge deletion.}
Remove $M\geq 1$ existing edges simultaneously, requiring at least one to be incident to $u$. All possible combinations of $M$ edges that satisfy this condition are considered, excluding modifications that disconnect the graph. The signatures $\tau_{G'}(u)$ of the resulting graphs are collected to construct $P(u)$.

\item \textbf{Node deletion.}
Remove the vertex $u$ along with all its incident edges. The probing profile $P(u)$ is constructed from the sorted signatures of all remaining vertices:
\[
P(u)=\operatorname{sort}\left(
\{\tau_{G'}(v):v\in V\setminus\{u\}\}
\right).
\]
\end{itemize}

\section{The $k$-dimensional Weisfeiler--Leman algorithm}

The Weisfeiler--Leman (WL) algorithm is a combinatorial method based on iterative color refinement, commonly used in graph isomorphism testing and vertex-orbit identification. Color refinement starts by assigning colors to vertices according to their degrees and repeatedly refines these colors based on the multisets of colors of their neighbors. The process continues until the coloring stabilizes, producing a partition of vertices into color classes. The $k$-dimensional Weisfeiler--Leman algorithm ($k$-WL), outlined in Algorithm~\ref{alg:k-wl}, generalizes this refinement procedure by assigning colors to ordered $k$-tuples of vertices. Color refinement corresponds to the one-dimensional variant (1-WL). Initially, each $k$-tuple is assigned a color based on its atomic type, defined as its equality pattern and the adjacency relations among its coordinates. At each iteration, the $k$-tuple is re-assigned a new color based on its previous color and the multiset of colors of the tuples obtained by replacing each of its coordinates, one at a time, with every vertex in the graph. For fixed $k$, the refinement maintains $N^k$ tuple colors and examines $O(kN^{k+1})$ coordinate replacements per iteration, before accounting for the cost of encoding and comparing colors.

\begin{algorithm}[H]
\caption{The $k$-dimensional Weisfeiler--Leman algorithm}
\label{alg:k-wl}
\begin{algorithmic}[1]
\Require An uncolored, undirected graph $\mathcal{G}=(V,E)$;
         dimension $k \geq 1$.
\Ensure The stable coloring of $V^k$.

\State \textbf{Initialization:}
\ForAll{tuples $\bar{v} \in V^k$}
    \State $C[\bar{v}] \gets \operatorname{atp}_k(\bar{v})$
\EndFor

\Repeat
    \ForAll{tuples $\bar{v}=(v_1,\ldots,v_k)\in V^k$}
        \State $M \gets \emptyset$
        \ForAll{vertices $u \in V$}
            \For{$i=1$ to $k$}
                \State $\bar{w} \gets
                (v_1,\ldots,v_{i-1},u,v_{i+1},
                \ldots,v_k)$
                \State $c_i \gets C[\bar{w}]$
            \EndFor
            \State Add $(\operatorname{atp}_{k+1}
            (v_1,\ldots,v_k,u),c_1,\ldots,c_k)$
            to $M$
        \EndFor
        \State $C_{\mathrm{new}}[\bar{v}]
        \gets (C[\bar{v}],M)$
    \EndFor
    \State $C \gets C_{\mathrm{new}}$
\Until{the coloring stabilizes}

\State \Return $C$
\end{algorithmic}
\end{algorithm}

An optimized implementation computes the stabilized $k$-WL coloring in time $O(k^2N^{k+1}\log N)$, which is $O(N^{k+1}\log N)$ for fixed $k$~\cite{ImmermanSengupta2019}. This bound covers the entire refinement until stabilization, rather than a single iteration. For 1-WL, the tighter bound $O((N+M)\log N)$ is known, where $M$ is the number of edges~\cite{BerkholzBonsmaGrohe2017}.

\begin{figure}[H]
    \centering
    \begin{subfigure}[b]{0.45\columnwidth}
        \centering
        \includegraphics[width=\linewidth]{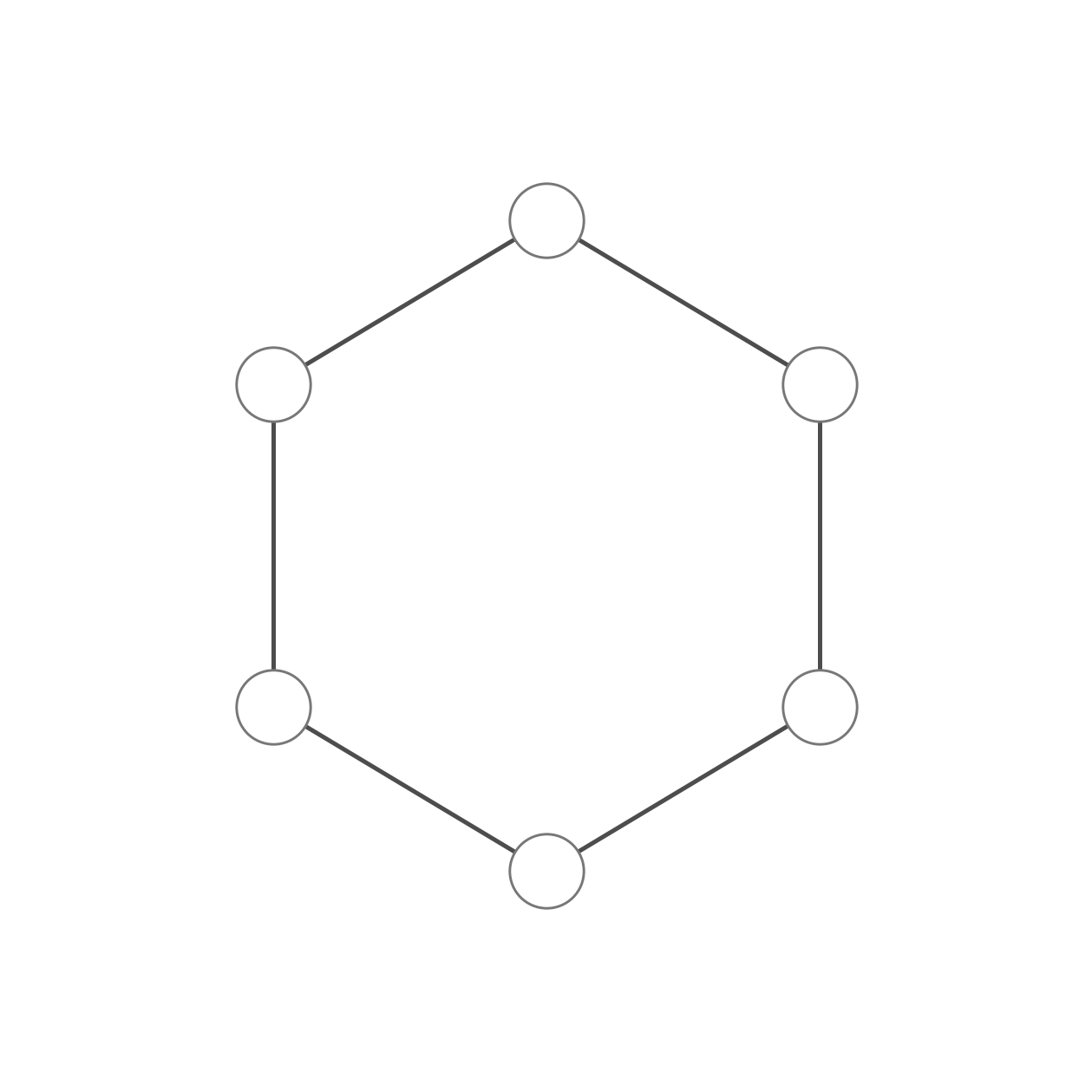}
        \caption{$C_6$}
    \end{subfigure}
    \hfill
    \begin{subfigure}[b]{0.45\columnwidth}
        \centering
        \includegraphics[width=\linewidth]{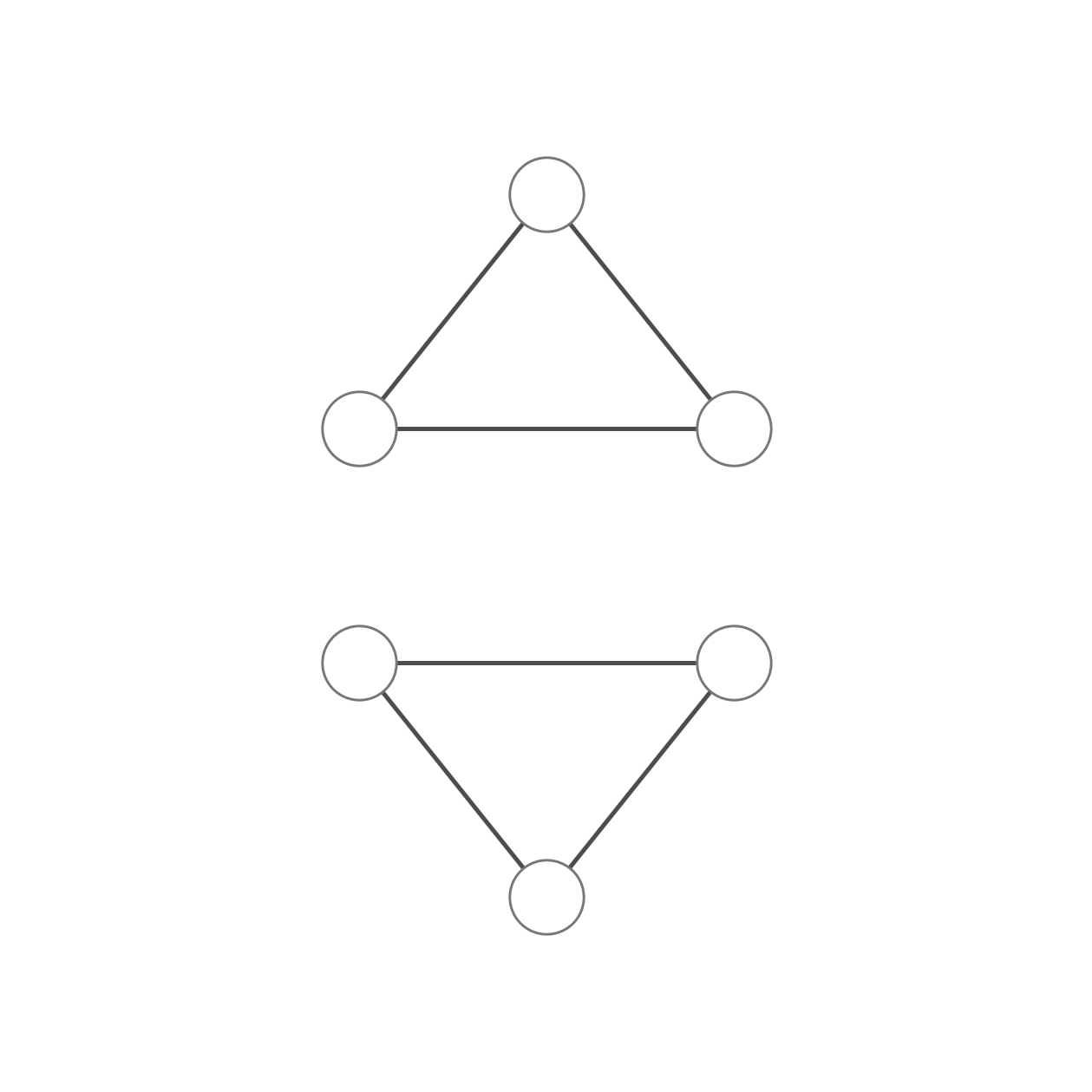}
        \caption{$2K_3$}
    \end{subfigure}
    \caption{Two non-isomorphic $2$-regular graphs that cannot be distinguished by 1-WL.}
    \label{fig:wl-original}
\end{figure}

\begin{figure}[H]
    \centering
    \begin{subfigure}[b]{0.45\columnwidth}
        \centering
        \includegraphics[width=\linewidth]{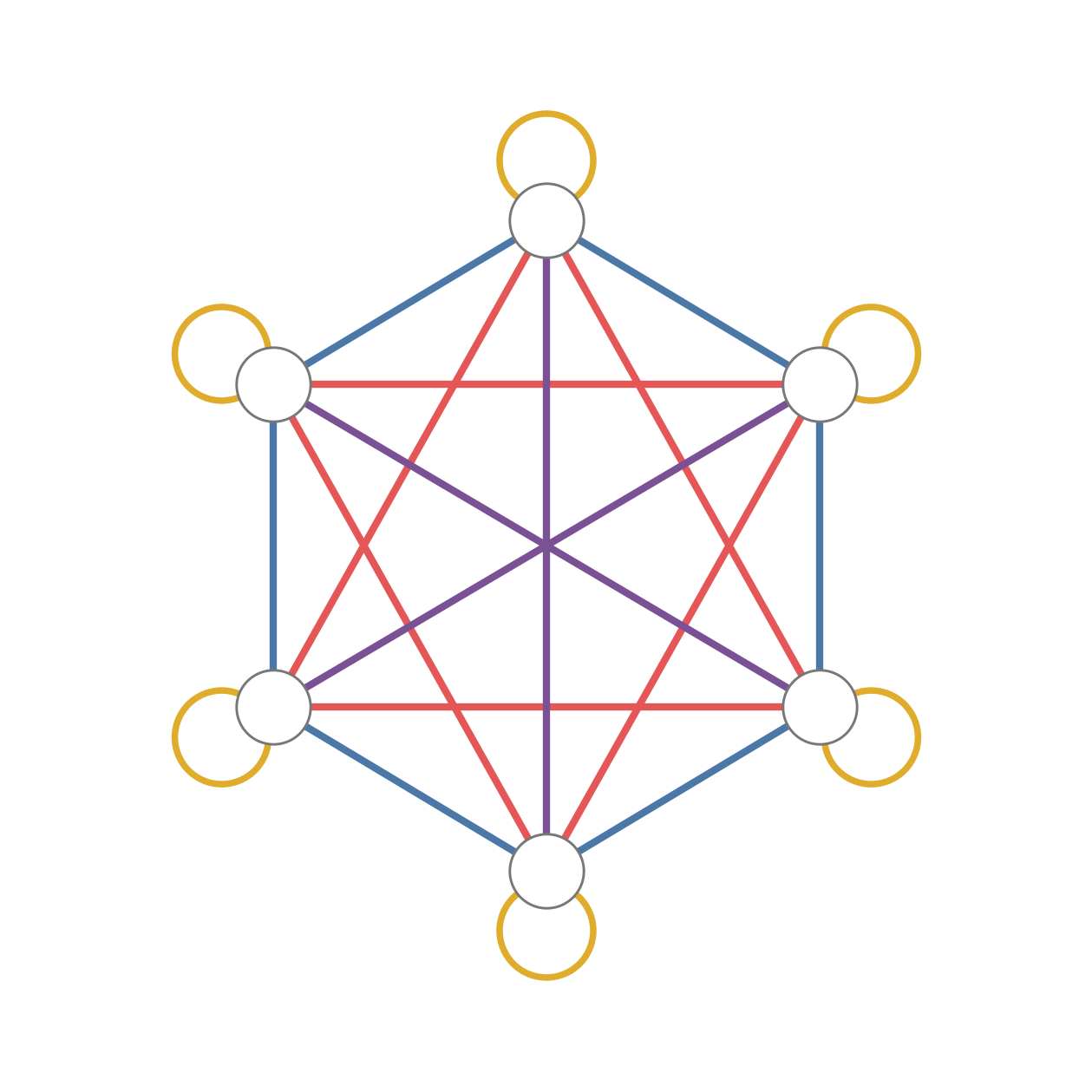}
        \caption{$C_6$}
    \end{subfigure}
    \hfill
    \begin{subfigure}[b]{0.45\columnwidth}
        \centering
        \includegraphics[width=\linewidth]{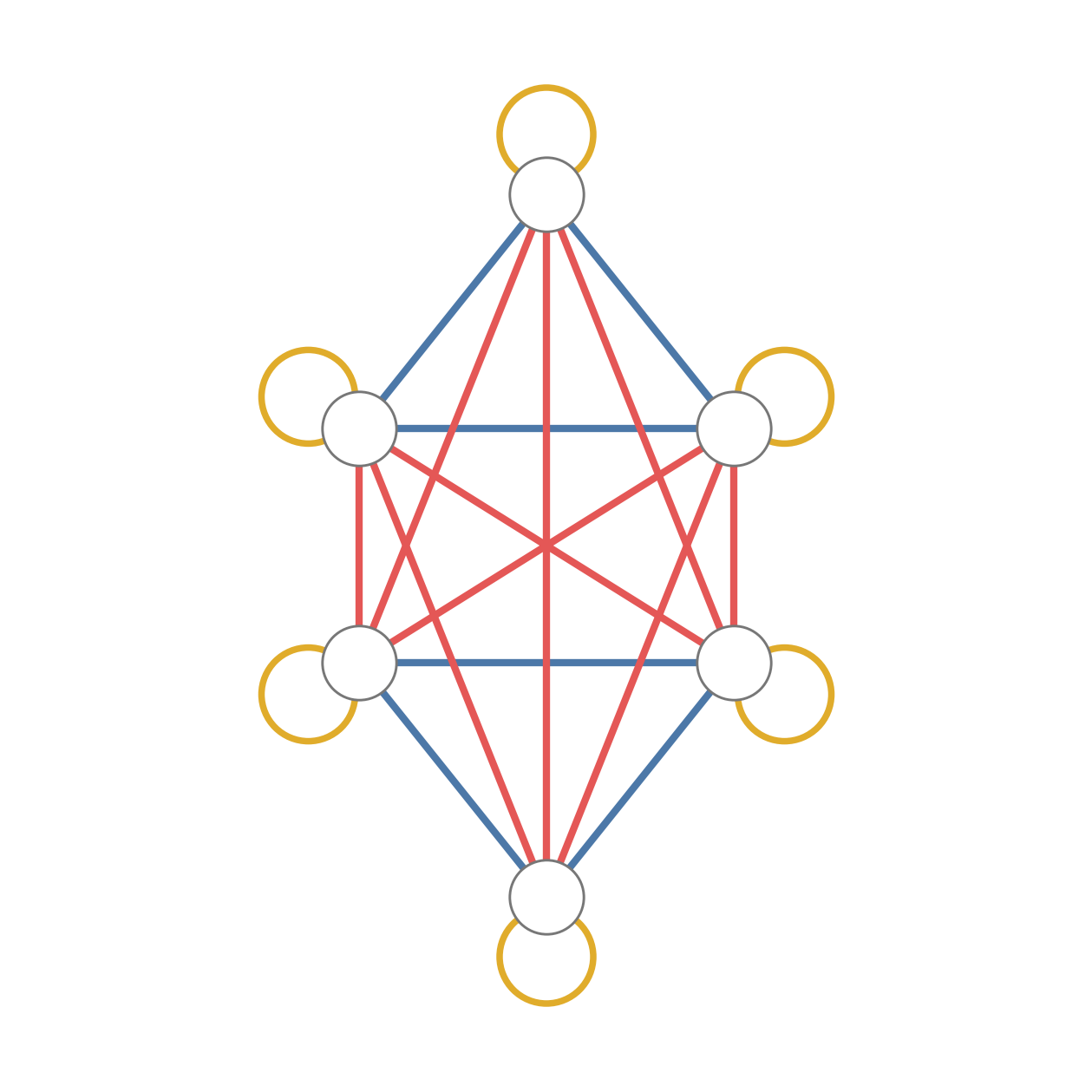}
        \caption{$2K_3$}
    \end{subfigure}
    \caption{The 2-WL colorings of $C_6$ and $2K_3$.}
    \label{fig:wl-2}
\end{figure}

Figure~\ref{fig:wl-original} illustrates two non-isomorphic graphs: the cycle $C_6$ and the disjoint union of two triangles, $2K_3$. Since both graphs are $2$-regular, color refinement (1-WL) assigns the same color to all vertices in both graphs and therefore fails to distinguish them as graphs. As shown in Figure~\ref{fig:wl-2}, 2-WL does distinguish the two graphs. This example concerns graph distinction rather than orbit recovery: each graph is vertex-transitive, so 1-WL already recovers its single vertex orbit. Our experiments instead compare the vertex partitions induced by $k$-WL directly with the true orbit partitions.

For vertex-orbit identification, the stabilized $k$-tuple coloring induces a vertex coloring by assigning to each vertex $v$ the color of its diagonal tuple $(v,\ldots,v)$. Vertices with the same induced color are placed in the same class. Because graph automorphisms preserve WL colors, each resulting class is a union of vertex orbits; nevertheless, distinct orbits may remain in the same class. We say that $k$-WL recovers the orbit partition of $\mathcal{G}$ when its induced vertex partition equals the partition into automorphism orbits. 

\section{Detailed experimental results}

\subsection{F-LEX Graphs}

\begin{table}[ht]
\centering
\caption{Results on F-LEX graphs under fixed-resolution comparison ($r=10^{9}$).}
\label{tab:flex-results}
\scriptsize
\begin{tabular}{lrccccc}
\toprule
\textbf{Graph} & $\mathbf{n}$ & \textbf{Method} &
\textbf{NMI} & $\mathbf{P}$ & \textbf{Classes} &
\textbf{$k$-WL} \\
\midrule

\multirow{2}{*}{f-lex-reg-10-1}
& \multirow{2}{*}{340}
& $G$-Laplacian
& 0.977331 & 0.885111 & 141 & -- \\

& & \textbf{$G$-Neighborhood deletion ($p=3$, $d=4$)}
& \textbf{1.000000} & \textbf{1.000000} & \textbf{175} & \\
\midrule
\multirow{2}{*}{f-lex-reg-10-3}
& \multirow{2}{*}{340}
& $G$-Laplacian
& 0.985687 & 0.925230 & 163 & -- \\

& & \textbf{$G$-Neighborhood deletion ($p=3$)}
& \textbf{1.000000} & \textbf{1.000000} & \textbf{184} & \\

\bottomrule
\end{tabular}
\end{table}

\subsection{USR Graphs}
\begin{longtable}{ll l c c c c}
\caption{Results for the USR graphs.}
\label{tab:usr}\\
\toprule
Graph & $n$ & Method & NMI & $P$ & Classes & $k$-WL \\
\midrule
\endfirsthead

\toprule
Graph & $n$ & Method & NMI & $P$ & Classes & $k$-WL \\
\midrule
\endhead

\multirow{3}{*}{\texttt{usr(1)\_29-1}}
& \multirow{3}{*}{29} & $G$-Laplacian & 0.000000 & 0.091558 & 1 & 4 \\
& & $G$-Correlation & 0.000000 & 0.091558 & 1 & \\
& & \textbf{$G$-Dirac} & \textbf{1.000000} & \textbf{1.000000} & \textbf{13} & \\
\midrule

\multirow{3}{*}{\texttt{usr(1)\_29-2}}
& \multirow{3}{*}{29} & $G$-Laplacian & 0.000000 & 0.034483 & 1 & 4 \\
& & $G$-Correlation & 0.000000 & 0.034483 & 1 & \\
& & \textbf{$G$-Dirac} & \textbf{1.000000} & \textbf{1.000000} & \textbf{29} & \\
\midrule

\multirow{3}{*}{\texttt{usr(2)\_58-1}}
& \multirow{3}{*}{58} & $G$-Laplacian & 0.000000 & 0.038644 & 1 & 4 \\
& & $G$-Correlation & 0.000000 & 0.038644 & 1 & \\
\bottomrule
\end{longtable}
\subsection{Had-SW Graphs}
\scriptsize
\begin{longtable}{l r c c c c c c}
\caption{Results on Had-SW graphs.}
\label{tab:had-sw-results}\\
\toprule
\textbf{Graph} & \textbf{$n$} & \textbf{Method} & \textbf{Resolution} & \textbf{NMI} & \textbf{$P$} & \textbf{Classes} & \textbf{$k$-WL} \\
\midrule
\endfirsthead
\toprule
\textbf{Graph} & \textbf{$n$} & \textbf{Method} & \textbf{Resolution} & \textbf{NMI} & \textbf{$P$} & \textbf{Classes} & \textbf{$k$-WL} \\
\midrule
\endhead
\texttt{had-sw-20} & 80 & \textbf{$G$-Laplacian} & -- & \textbf{1.000000} & \textbf{1.000000} & \textbf{1} & 1 \\
\midrule
\texttt{had-sw-32-1} & 128 & $G$-Laplacian & -- & 0.000000 & 0.026367 & 1 & $>3$ \\
 &  & $G$-Correlation & -- & 0.000000 & 0.026367 & 1 &  \\
 &  & $G$-Dirac & -- & 0.000000 & 0.026367 & 1 &  \\
 &  & $G$-Node deletion ($M=1$) & -- & 0.000000 & 0.026367 & 1 &  \\
 &  & \textbf{$G$-Neighborhood deletion ($p=3$)} & -- & \textbf{1.000000} & \textbf{1.000000} & \textbf{42} &  \\
\midrule
\texttt{had-sw-32-2} & 128 & $G$-Laplacian & -- & 0.000000 & 0.027344 & 1 & $>3$ \\
 &  & $G$-Correlation & -- & 0.000000 & 0.027344 & 1 &  \\
\bottomrule
\end{longtable}

\subsection{Grid Graphs*}

\begin{longtable}{l r c c c c c c}
\caption{Results on Grid graphs with adaptive fixed-resolution comparison.}
\label{tab:grid-results}\\

\toprule
\textbf{Graph} & \textbf{$n$} & \textbf{Method} & \textbf{Resolution} &
\textbf{NMI} & \textbf{$P$} & \textbf{Classes} & \textbf{$k$-WL} \\
\midrule
\endfirsthead

\toprule
\textbf{Graph} & \textbf{$n$} & \textbf{Method} & \textbf{Resolution} &
\textbf{NMI} & \textbf{$P$} & \textbf{Classes} & \textbf{$k$-WL} \\
\midrule
\endhead

grid-2-5  & 25   & \textbf{G-Laplacian} & $10^{8}$ & \textbf{1.000000} & \textbf{1.000000} & \textbf{6}   & 1 \\
\midrule
grid-2-10 & 100  & \textbf{G-Laplacian} & $10^{8}$ & \textbf{1.000000} & \textbf{1.000000} & \textbf{15}  & 1 \\
\midrule
grid-2-15 & 225  & \textbf{G-Laplacian} & $10^{8}$ & \textbf{1.000000} & \textbf{1.000000} & \textbf{36}  & 1 \\
\midrule
grid-2-20 & 400  & \textbf{G-Laplacian} & $10^{8}$ & \textbf{1.000000} & \textbf{1.000000} & \textbf{55}  & 1 \\
\midrule
grid-2-25 & 625  & \textbf{G-Laplacian} & $10^{3}$ & \textbf{1.000000} & \textbf{1.000000} & \textbf{91}  & 1 \\
\midrule
grid-2-30 & 900  & \textbf{G-Laplacian} & $10^{8}$ & \textbf{1.000000} & \textbf{1.000000} & \textbf{120} & 1 \\
\midrule
grid-2-35 & 1225 & \textbf{G-Laplacian} & $10^{8}$ & \textbf{1.000000} & \textbf{1.000000} & \textbf{171} & 1 \\
\midrule
grid-2-40 & 1600 & \textbf{G-Laplacian} & $10^{8}$ & \textbf{1.000000} & \textbf{1.000000} & \textbf{210} & 1 \\
\bottomrule
\end{longtable}

\subsection{Grid-W Graphs*}

\begin{longtable}{l r c c c c c c}
\caption{Results on Grid-W graphs with adaptive fixed-resolution comparison.}
\label{tab:grid-w-results}\\

\toprule
\textbf{Graph} & \textbf{$n$} & \textbf{Method} & \textbf{Resolution} &
\textbf{NMI} & \textbf{$P$} & \textbf{Classes} & \textbf{$k$-WL} \\
\midrule
\endfirsthead

\toprule
\textbf{Graph} & \textbf{$n$} & \textbf{Method} & \textbf{Resolution} &
\textbf{NMI} & \textbf{$P$} & \textbf{Classes} & \textbf{$k$-WL} \\
\midrule
\endhead

grid-w-2-5  & 25   & \textbf{G-Laplacian} & $10^{8}$ & \textbf{1.000000} & \textbf{1.000000} & \textbf{1} & 1 \\
\midrule
grid-w-2-10 & 100  & \textbf{G-Laplacian} & $10^{8}$ & \textbf{1.000000} & \textbf{1.000000} & \textbf{1} & 1 \\
\midrule
grid-w-2-15 & 225  & \textbf{G-Laplacian} & $10^{8}$ & \textbf{1.000000} & \textbf{1.000000} & \textbf{1} & 1 \\
\midrule
grid-w-2-20 & 400  & \textbf{G-Laplacian} & $10^{8}$ & \textbf{1.000000} & \textbf{1.000000} & \textbf{1} & 1 \\
\midrule
grid-w-2-25 & 625  & \textbf{G-Laplacian} & $10^{8}$ & \textbf{1.000000} & \textbf{1.000000} & \textbf{1} & 1 \\
\midrule
grid-w-2-30 & 900  & \textbf{G-Laplacian} & $10^{8}$ & \textbf{1.000000} & \textbf{1.000000} & \textbf{1} & 1 \\
\midrule
grid-w-2-35 & 1225 & \textbf{G-Laplacian} & $10^{8}$ & \textbf{1.000000} & \textbf{1.000000} & \textbf{1} & 1 \\
\midrule
grid-w-2-40 & 1600 & \textbf{G-Laplacian} & $10^{8}$ & \textbf{1.000000} & \textbf{1.000000} & \textbf{1} & 1 \\
\bottomrule
\end{longtable}

\subsection{dac Graphs }

\begin{longtable}{@{}l r c c c c c c@{}}
\caption{Results on dac graphs with adaptive fixed-resolution comparison.}
\label{tab:bliss-results}\\

\toprule
\textbf{Graph} & \textbf{$n$} & \textbf{Method} & \textbf{Resolution} &
\textbf{NMI} & \textbf{$P$} & \textbf{Classes} & \textbf{$k$-WL} \\
\midrule
\endfirsthead

\toprule
\textbf{Graph} & \textbf{$n$} & \textbf{Method} & \textbf{Resolution} &
\textbf{NMI} & \textbf{$P$} & \textbf{Classes} & \textbf{$k$-WL} \\
\midrule
\endhead

2pipe.bliss & 3575 & \textbf{G-Laplacian} & $10^{8}$ &
\textbf{1.000000} & \textbf{1.000000} & \textbf{1918} & 1 \\
\midrule
3pipe.bliss & 10048 & \textbf{G-Laplacian} & $10^{8}$ &
\textbf{1.000000} & \textbf{1.000000} & \textbf{5090} & 1 \\
\midrule
fpga11\_13.bliss & 598 & \textbf{G-Laplacian} & $10^{8}$ &
\textbf{1.000000} & \textbf{1.000000} & \textbf{3} & 1 \\
\midrule
fpga11\_20.bliss & 920 & \textbf{G-Laplacian} & $10^{8}$ &
\textbf{1.000000} & \textbf{1.000000} & \textbf{3} & 1 \\
\midrule
fpga13\_10.bliss & 530 & \textbf{G-Laplacian} & $10^{8}$ &
\textbf{1.000000} & \textbf{1.000000} & \textbf{11} & 1 \\
\midrule
fpga13\_12.bliss & 636 & \textbf{G-Laplacian} & $10^{8}$ &
\textbf{1.000000} & \textbf{1.000000} & \textbf{11} & 1 \\
\midrule
hole11.bliss & 276 & \textbf{G-Laplacian} & $10^{8}$ &
\textbf{1.000000} & \textbf{1.000000} & \textbf{3} & 1 \\
\midrule
hole12.bliss & 325 & \textbf{G-Laplacian} & $10^{8}$ &
\textbf{1.000000} & \textbf{1.000000} & \textbf{3} & 1 \\
\bottomrule
\end{longtable}

\subsection{Hypercube Graphs}

\begin{longtable}{l r c c c c c}
\caption{Results on Hypercube graphs.}
\label{tab:hypercube-results}\\

\toprule
\textbf{Graph} & \textbf{$n$} & \textbf{Method} &
\textbf{NMI} & \textbf{$P$} & \textbf{Classes} & \textbf{$k$-WL} \\
\midrule
\endfirsthead

\toprule
\textbf{Graph} & \textbf{$n$} & \textbf{Method} &
\textbf{NMI} & \textbf{$P$} & \textbf{Classes} & \textbf{$k$-WL} \\
\midrule
\endhead

3cube.bliss  & 8    & \textbf{G-Laplacian} & \textbf{1.000000} & \textbf{1.000000} & \textbf{1} & 1 \\
4cube.bliss  & 16   & \textbf{G-Laplacian} & \textbf{1.000000} & \textbf{1.000000} & \textbf{1} & 1 \\
5cube.bliss  & 32   & \textbf{G-Laplacian} & \textbf{1.000000} & \textbf{1.000000} & \textbf{1} & 1 \\
6cube.bliss  & 64   & \textbf{G-Laplacian} & \textbf{1.000000} & \textbf{1.000000} & \textbf{1} & 1 \\
7cube.bliss  & 128  & \textbf{G-Laplacian} & \textbf{1.000000} & \textbf{1.000000} & \textbf{1} & 1 \\
8cube.bliss  & 256  & \textbf{G-Laplacian} & \textbf{1.000000} & \textbf{1.000000} & \textbf{1} & 1 \\
9cube.bliss  & 512  & \textbf{G-Laplacian} & \textbf{1.000000} & \textbf{1.000000} & \textbf{1} & 1 \\
10cube.bliss & 1024 & \textbf{G-Laplacian} & \textbf{1.000000} & \textbf{1.000000} & \textbf{1} & 1 \\
11cube.bliss & 2048 & \textbf{G-Laplacian} & \textbf{1.000000} & \textbf{1.000000} & \textbf{1} & 1 \\
12cube.bliss & 4096 & \textbf{G-Laplacian} & \textbf{1.000000} & \textbf{1.000000} & \textbf{1} & 1 \\
13cube.bliss & 8192 & \textbf{G-Laplacian} & \textbf{1.000000} & \textbf{1.000000} & \textbf{1} & 1 \\ 
14cube.bliss & 16384 & \textbf{G-Laplacian} & \textbf{1.000000} & \textbf{1.000000} & \textbf{1} & 1 \\ 
15cube.bliss & 32768 & \textbf{G-Laplacian} & \textbf{1.000000} & \textbf{1.000000} & \textbf{1} & 1 \\
16cube.bliss & 65536 & \textbf{$G$-Laplacian} & \textbf{1.000000} & \textbf{1.000000} & \textbf{1} & 1 \\
\bottomrule
\end{longtable}

\subsection{Latin-SW Graphs}

\begin{longtable}{l r c c c c c}
\caption{Results on Latin-SW graphs.}
\label{tab:latin-sw-results}\\

\toprule
\textbf{Graph} & \textbf{$n$} & \textbf{Method} &
\textbf{NMI} & \textbf{$P$} & \textbf{Classes} & \textbf{$k$-WL} \\
\midrule
\endfirsthead

\toprule
\textbf{Graph} & \textbf{$n$} & \textbf{Method} &
\textbf{NMI} & \textbf{$P$} & \textbf{Classes} & \textbf{$k$-WL} \\
\midrule
\endhead

\multirow{3}{*}{\texttt{latin-sw-10-1}}
& \multirow{3}{*}{100} & $G$-Laplacian & 0.000000 & 0.010000 & 1 & $>$3 \\
& & $G$-Correlation & 0.706389 & 0.255820 & 18 &  \\
& & \textbf{$G$-Dirac} & \textbf{1.000000} & \textbf{1.000000} & \textbf{100} &  \\
\midrule

\multirow{6}{*}{\texttt{latin-sw-10-2}}
& \multirow{6}{*}{100} & $G$-Laplacian & 0.000000 & 0.010000 & 1 & -- \\
& & $G$-Correlation & 0.753864 & 0.224623 & 20 &  \\
& & $G$-Dirac & 0.998493 & 0.500000 & 99 &  \\
& & $G_2$-Laplacian & 0.998493 & 0.500000 & 99 &  \\
& & $G_2$-Correlation & 0.998493 & 0.500000 & 99 &  \\
\bottomrule
\end{longtable}

\subsection{Hadamard Graphs*}

\begin{longtable}{l r c c c c c c}
\caption{Results on had graphs.}
\label{tab:had-results}\\

\toprule
\textbf{Graph} & \textbf{$n$} & \textbf{Method} & \textbf{Resolution} &
\textbf{NMI} & \textbf{$P$} & \textbf{Classes} & \textbf{$k$-WL} \\
\midrule
\endfirsthead

\toprule
\textbf{Graph} & \textbf{$n$} & \textbf{Method} & \textbf{Resolution} &
\textbf{NMI} & \textbf{$P$} & \textbf{Classes} & \textbf{$k$-WL} \\
\midrule
\endhead

had-1   & 4   & \textbf{G-Laplacian} & $10^{8}$ & \textbf{1.000000} & \textbf{1.000000} & \textbf{1} & 1 \\
had-2   & 8   & \textbf{G-Laplacian} & $10^{8}$ & \textbf{1.000000} & \textbf{1.000000} & \textbf{1} & 1 \\
had-4   & 16  & \textbf{G-Laplacian} & $10^{8}$ & \textbf{1.000000} & \textbf{1.000000} & \textbf{1} & 1 \\
had-8   & 32  & \textbf{G-Laplacian} & $10^{8}$ & \textbf{1.000000} & \textbf{1.000000} & \textbf{1} & 1 \\
had-12  & 48  & \textbf{G-Laplacian} & $10^{8}$ & \textbf{1.000000} & \textbf{1.000000} & \textbf{1} & 1 \\
had-16  & 64  & \textbf{G-Laplacian} & $10^{8}$ & \textbf{1.000000} & \textbf{1.000000} & \textbf{1} & 1 \\
had-20  & 80  & \textbf{G-Laplacian} & $10^{8}$ & \textbf{1.000000} & \textbf{1.000000} & \textbf{1} & 1 \\
had-24  & 96  & \textbf{G-Laplacian} & $10^{8}$ & \textbf{1.000000} & \textbf{1.000000} & \textbf{1} & 1 \\
\bottomrule
\end{longtable}

\subsection{MZ Graphs*}

\begin{longtable}{l r c c c c c c}
\caption{Results on MZ graphs with adaptive fixed-resolution comparison.}
\label{tab:mz-results}\\

\toprule
\textbf{Graph} & \textbf{$n$} & \textbf{Method} & \textbf{Resolution} &
\textbf{NMI} & \textbf{$P$} & \textbf{Classes} & \textbf{$k$-WL} \\
\midrule
\endfirsthead

\toprule
\textbf{Graph} & \textbf{$n$} & \textbf{Method} & \textbf{Resolution} &
\textbf{NMI} & \textbf{$P$} & \textbf{Classes} & \textbf{$k$-WL} \\
\midrule
\endhead

mz-2 & 40 & \textbf{G-Laplacian} & $10^{8}$ &
\textbf{1.000000} & \textbf{1.000000} & \textbf{10} & 3 \\

\midrule

\multirow{7}{*}{mz-4} & \multirow{7}{*}{80}
& G-Laplacian     & $10^{8}$ & 0.962235 & 0.875000 & 16 & 3 \\
& & G-Correlation  & $10^{1}$ & 0.962235 & 0.875000 & 16 &  \\
& & G-Dirac        & $10^{8}$ & 0.962235 & 0.875000 & 16 &  \\
& & $G_2$-Laplacian    & $10^{8}$ & 0.962235 & 0.875000 & 16 &  \\
& & $G_2$-Correlation  & $10^{8}$ & 0.962235 & 0.875000 & 16 &  \\
& & $G_2$-Dirac        & $10^{8}$ & 0.962235 & 0.875000 & 16 &  \\
& & \textbf{$G_3$-Laplacian} & $10^{8}$ &
\textbf{1.000000} & \textbf{1.000000} & \textbf{20} &  \\
\bottomrule
\end{longtable}

\subsection{mz-aug2 Graphs*}

\begin{longtable}{l r c c c c c}
\caption{Results on mz-aug2 graphs under fixed-resolution comparison ($r=10^{6}$).}
\label{tab:mz-aug2-results}\\
\toprule
\textbf{Graph} & \textbf{$n$} & \textbf{Method} & \textbf{NMI} & \textbf{$P$} & \textbf{Classes} & \textbf{$k$-WL} \\
\midrule
\endfirsthead

\toprule
\textbf{Graph} & \textbf{$n$} & \textbf{Method} & \textbf{NMI} & \textbf{$P$} & \textbf{Classes} & \textbf{$k$-WL} \\
\midrule
\endhead

\multirow{7}{*}{\texttt{mz-aug2-4}}
& \multirow{7}{*}{96} & $G$-Laplacian     & 0.969240 & 0.785714 & 44 & 3 \\
& & $G$-Correlation   & 0.969240 & 0.785714 & 44 &  \\
& & $G$-Dirac         & 0.969240 & 0.785714 & 44 &  \\
& & $G_2$-Laplacian   & 0.977106 & 0.833333 & 46 &  \\
& & $G_2$-Correlation & 0.977106 & 0.833333 & 46 &  \\
& & $G_2$-Dirac       & 0.977106 & 0.833333 & 46 &  \\
& & \textbf{$G_3$-Laplacian} & \textbf{1.000000} & \textbf{1.000000} & \textbf{56} &  \\
\midrule

\multirow{7}{*}{\texttt{mz-aug2-6}}
& \multirow{7}{*}{144} & $G$-Laplacian     & 0.962637 & 0.727273 & 60 & 3 \\
\bottomrule
\end{longtable}

\subsection{mz-aug Graphs}

\begin{longtable}{l r c c c c c c}
\caption{Results on mz-aug graphs.}
\label{tab:mz-aug-results}\\
\toprule
\textbf{Graph} & \textbf{$n$} & \textbf{Method} &
\textbf{Resolution} & \textbf{NMI} & \textbf{$P$} &
\textbf{Classes} & \textbf{$k$-WL} \\
\midrule
\endfirsthead

\toprule
\textbf{Graph} & \textbf{$n$} & \textbf{Method} &
\textbf{Resolution} & \textbf{NMI} & \textbf{$P$} &
\textbf{Classes} & \textbf{$k$-WL} \\
\midrule
\endhead

\texttt{mz-aug-2} & 40 &
\textbf{$G$-Laplacian} &  $10^{9}$  &
\textbf{1.000000} & \textbf{1.000000} & \textbf{10} & 3 \\
\midrule

\texttt{mz-aug-4} & 80 &
$G$-Laplacian &  $10^{9}$  &
0.962235 & 0.875000 & 16 & 3 \\

& & $G$-Correlation & $10^{1}$ &
0.962235 & 0.875000 & 16 &  \\

& & $G$-Dirac &  $10^{9}$  &
0.962235 & 0.875000 & 16 &  \\

& & \textbf{$G$-Node deletion} &  $10^{9}$  &
\textbf{1.000000} & \textbf{1.000000} & \textbf{20} &  \\
\midrule

\texttt{mz-aug-6} & 120 &
$G$-Laplacian &  $10^{9}$  &
0.955624 & 0.818182 & 22 & 3 \\

& & $G$-Correlation & $10^{1}$ &
0.955624 & 0.818182 & 22 &  \\

& & $G$-Dirac &  $10^{9}$  &
0.955624 & 0.818182 & 22 &  \\
\bottomrule
\end{longtable}

\subsection{CHH Graphs}

\begin{longtable}{l r c c c c c}
\caption{Results on CHH graphs under fixed-resolution comparison ($r=10^{6}$).}
\label{tab:chh-results}\\
\toprule
\textbf{Graph} & \textbf{$n$} & \textbf{Method} & \textbf{NMI} & \textbf{$P$} & \textbf{Classes} & \textbf{$k$-WL} \\
\midrule
\endfirsthead

\toprule
\textbf{Graph} & \textbf{$n$} & \textbf{Method} & \textbf{NMI} & \textbf{$P$} & \textbf{Classes} & \textbf{$k$-WL} \\
\midrule
\endhead

\texttt{CHH\_cc(1-1)\_22-1} & 22 & \textbf{$G$-Laplacian} & \textbf{1.000000} & \textbf{1.000000} & \textbf{4} & 3 \\
\midrule

\texttt{CHH\_cc(1-1)\_22-2} & 22 & \textbf{$G$-Laplacian} & \textbf{1.000000} & \textbf{1.000000} & \textbf{4} & 3 \\
\midrule

\texttt{CHH\_cc(2-1)\_44-1} & 44 & \textbf{$G$-Laplacian} & \textbf{1.000000} & \textbf{1.000000} & \textbf{8} & 3 \\
\midrule

\texttt{CHH\_cc(2-1)\_44-2} & 44 & \textbf{$G$-Laplacian} & \textbf{1.000000} & \textbf{1.000000} & \textbf{4} & 1 \\
\midrule

\texttt{CHH\_cc(2-2)\_88-1} & 88 & \textbf{$G$-Laplacian} & \textbf{1.000000} & \textbf{1.000000} & \textbf{8} & 3 \\
\midrule

\multirow{6}{*}{\texttt{CHH\_cc(2-2)\_88-2}}
& \multirow{6}{*}{88} & $G$-Laplacian     & 0.994897 & 0.955556 & 11 & 3 \\
& & $G$-Correlation   & 0.994897 & 0.955556 & 11 &  \\
& & $G$-Dirac         & 0.994897 & 0.955556 & 11 &  \\
\bottomrule
\end{longtable}

\subsection{PP Graphs*}

\begin{longtable}{l r p{3.5cm} c c c c}
\caption{Results on PP graphs under fixed-resolution comparison ($r=10^{6}$).}
\label{tab:pp-results}\\
\toprule
\textbf{Graph} & \textbf{$n$} & \textbf{Method} & \textbf{NMI} & \textbf{$P$} & \textbf{Classes} & \textbf{$k$-WL} \\
\midrule
\endfirsthead

\toprule
\textbf{Graph} & \textbf{$n$} & \textbf{Method} & \textbf{NMI} & \textbf{$P$} & \textbf{Classes} & \textbf{$k$-WL} \\
\midrule
\endhead

\texttt{pp-2-1} & 14 & \textbf{$G$-Laplacian} & \textbf{1.000000} & \textbf{1.000000} & \textbf{1} & 1 \\
\midrule

\texttt{pp-3-1} & 26 & \textbf{$G$-Laplacian} & \textbf{1.000000} & \textbf{1.000000} & \textbf{1} & 1 \\
\midrule

\texttt{pp-4-1} & 42 & \textbf{$G$-Laplacian} & \textbf{1.000000} & \textbf{1.000000} & \textbf{1} & 1 \\
\midrule

\texttt{pp-5-1} & 62 & \textbf{$G$-Laplacian} & \textbf{1.000000} & \textbf{1.000000} & \textbf{1} & 1 \\
\midrule

\texttt{pp-7-1} & 114 & \textbf{$G$-Laplacian} & \textbf{1.000000} & \textbf{1.000000} & \textbf{1} & 1 \\
\midrule

\texttt{pp-8-1} & 146 & \textbf{$G$-Laplacian} & \textbf{1.000000} & \textbf{1.000000} & \textbf{1} & 1 \\
\midrule

\texttt{pp-9-1} & 182 & \textbf{$G$-Laplacian} & \textbf{1.000000} & \textbf{1.000000} & \textbf{1} & 1 \\
\midrule

\multirow{15}{*}{\texttt{pp-9-2}}
& \multirow{15}{*}{182} & $G$-Laplacian             & 0.000000 & 0.445659 & 1 & $>$3 \\
\bottomrule
\end{longtable}

\subsection{STS Graphs*}

\begin{longtable}{l r c c c c c}
\caption{Results on STS graphs under fixed-resolution comparison ($r=10^{6}$).}
\label{tab:sts-results}\\
\toprule
\textbf{Graph} & \textbf{$n$} & \textbf{Method} & \textbf{NMI} & \textbf{$P$} & \textbf{Classes} & \textbf{$k$-WL} \\
\midrule
\endfirsthead

\toprule
\textbf{Graph} & \textbf{$n$} & \textbf{Method} & \textbf{NMI} & \textbf{$P$} & \textbf{Classes} & \textbf{$k$-WL} \\
\midrule
\endhead

\texttt{sts-7} & 7 & \textbf{$G$-Laplacian} & \textbf{1.000000} & \textbf{1.000000} & \textbf{1} & 1 \\
\midrule

\texttt{sts-9} & 12 & \textbf{$G$-Laplacian} & \textbf{1.000000} & \textbf{1.000000} & \textbf{1} & 1 \\
\midrule

\multirow{3}{*}{\texttt{sts-13}}
& \multirow{3}{*}{26} & $G$-Laplacian   & 0.000000 & 0.500000 & 1 & 4 \\
& & $G$-Correlation & 0.000000 & 0.500000 & 1 &  \\
& & \textbf{$G$-Dirac} & \textbf{1.000000} & \textbf{1.000000} & \textbf{2} &  \\
\midrule

\multirow{6}{*}{\texttt{sts-15}}
& \multirow{6}{*}{35} & $G$-Laplacian     & 0.000000 & 0.755102 & 1 & 4 \\
& & $G$-Correlation   & 0.000000 & 0.755102 & 1 &  \\
& & $G$-Dirac         & 0.000000 & 0.755102 & 1 &  \\
\bottomrule
\end{longtable}

\subsection{STS-SW Graphs}

\begin{longtable}{l r c c c c c}
\caption{Results on STS-SW graphs at resolution $10^{8}$.}
\label{tab:sts-sw-results}\\

\toprule
\textbf{Graph} & \textbf{$n$} & \textbf{Method} &
\textbf{NMI} & \textbf{$P$} & \textbf{Classes} & \textbf{$k$-WL} \\
\midrule
\endfirsthead

\toprule
\textbf{Graph} & \textbf{$n$} & \textbf{Method} &
\textbf{NMI} & \textbf{$P$} & \textbf{Classes} & \textbf{$k$-WL} \\
\midrule
\endhead

\multirow{6}{*}{sts-sw-19-1} & \multirow{6}{*}{57} & $G$-Laplacian & 0.000000 & 0.017544 & 1 & 4 \\
& & $G$-Correlation & 0.000000 & 0.017544 & 1 &  \\
& & $G$-Dirac & 0.974151 & 0.476190 & 49 &  \\
& & $G_{2}$-Laplacian & 0.974151 & 0.476190 & 49 &  \\
& & $G_{2}$-Correlation & 0.974151 & 0.476190 & 49 &  \\
& & \textbf{$G_{2}$-Dirac} & \textbf{1.000000} & \textbf{1.000000} & \textbf{57} &  \\
\midrule

\multirow{6}{*}{sts-sw-19-2} & \multirow{6}{*}{57} & $G$-Laplacian & 0.000000 & 0.017544 & 1 & 4 \\
& & $G$-Correlation & 0.000000 & 0.017544 & 1 &  \\
\bottomrule
\end{longtable}

\subsection{CMZ Graphs}

\begin{longtable}{lrlcccc}
\caption{Results on CMZ graphs under fixed-resolution comparison ($r=10^{6}$).}
\label{tab:cmz-results}\\

\toprule
\textbf{Graph} & $\mathbf{n}$ & \textbf{Method} & \textbf{NMI} & $\mathbf{P}$ & \textbf{\# Classes} & \textbf{$k$-WL} \\
\midrule
\endfirsthead

\toprule
\textbf{Graph} & $\mathbf{n}$ & \textbf{Method} & \textbf{NMI} & $\mathbf{P}$ & \textbf{\# Classes} & \textbf{$k$-WL} \\
\midrule
\endhead

\bottomrule
\endfoot

cmz-10 & 240 & \textbf{$G$-Laplacian} & \textbf{1.000000} & \textbf{1.000000} & \textbf{8} & 1 \\
cmz-11 & 264 & \textbf{$G$-Laplacian} & \textbf{1.000000} & \textbf{1.000000} & \textbf{8} & 1 \\
cmz-12 & 288 & \textbf{$G$-Laplacian} & \textbf{1.000000} & \textbf{1.000000} & \textbf{8} & 1 \\
cmz-13 & 312 & \textbf{$G$-Laplacian} & \textbf{1.000000} & \textbf{1.000000} & \textbf{8} & 1 \\
cmz-14 & 336 & \textbf{$G$-Laplacian} & \textbf{1.000000} & \textbf{1.000000} & \textbf{8} & 1 \\
cmz-15 & 360 & \textbf{$G$-Laplacian} & \textbf{1.000000} & \textbf{1.000000} & \textbf{8} & 1 \\
cmz-16 & 384 & \textbf{$G$-Laplacian} & \textbf{1.000000} & \textbf{1.000000} & \textbf{8} & 1 \\
cmz-17 & 408 & \textbf{$G$-Laplacian} & \textbf{1.000000} & \textbf{1.000000} & \textbf{8} & 1 \\
\end{longtable}

\subsection{SAT-CFI Graphs}

\begin{longtable}{l r c c c c c}
\caption{Results on sat\_cfi\_dim graphs at resolution $10^{8}$.}
\label{tab:sat-cfi-results}\\

\toprule
\textbf{Graph} & \textbf{$n$} & \textbf{Method} &
\textbf{NMI} & \textbf{$P$} & \textbf{Classes} & \textbf{$k$-WL} \\
\midrule
\endfirsthead

\toprule
\textbf{Graph} & \textbf{$n$} & \textbf{Method} &
\textbf{NMI} & \textbf{$P$} & \textbf{Classes} & \textbf{$k$-WL} \\
\midrule
\endhead

\texttt{sat\_cfi\_base\_0100\_a.dmc} & 30 & \textbf{$G$-Laplacian} & \textbf{1.000000} & \textbf{1.000000} & \textbf{30} & 1 \\
\midrule

\texttt{sat\_cfi\_base\_0100\_b.dmc} & 30 & \textbf{$G$-Laplacian} & \textbf{1.000000} & \textbf{1.000000} & \textbf{30} & 1 \\
\midrule

\texttt{sat\_cfi\_base\_0200\_a.dmc} & 60 & \textbf{$G$-Laplacian} & \textbf{1.000000} & \textbf{1.000000} & \textbf{60} & 1 \\
\midrule

\texttt{sat\_cfi\_base\_0200\_b.dmc} & 60 & \textbf{$G$-Laplacian} & \textbf{1.000000} & \textbf{1.000000} & \textbf{60} & 1 \\
\midrule

\texttt{sat\_cfi\_base\_0300\_a.dmc} & 90 & \textbf{$G$-Laplacian} & \textbf{1.000000} & \textbf{1.000000} & \textbf{90} & 1 \\
\midrule

\texttt{sat\_cfi\_base\_0300\_b.dmc} & 90 & \textbf{$G$-Laplacian} & \textbf{1.000000} & \textbf{1.000000} & \textbf{90} & 1 \\
\midrule

\texttt{sat\_cfi\_base\_0400\_a.dmc} & 120 & \textbf{$G$-Laplacian} & \textbf{1.000000} & \textbf{1.000000} & \textbf{120} & 1 \\
\midrule

\texttt{sat\_cfi\_base\_0400\_b.dmc} & 120 & \textbf{$G$-Laplacian} & \textbf{1.000000} & \textbf{1.000000} & \textbf{120} & 1 \\
\bottomrule
\end{longtable}

\subsection{CFI Graphs}
\begin{longtable}{lcccc c}
\caption{Results on CFI graphs under fixed-resolution comparison ($r=10^{6}$).}
\label{tab:cfi-results}\\
\toprule
\textbf{Graph} & \textbf{Method} & \textbf{NMI} & $\mathbf{P}$ & \textbf{\# Classes} & \textbf{$k$-WL} \\
\midrule
\endfirsthead

\toprule
\textbf{Graph} & \textbf{Method} & \textbf{NMI} & $\mathbf{P}$ & \textbf{\# Classes} & \textbf{$k$-WL} \\
\midrule
\endhead

cfi-20  & \textbf{Laplacian} & \textbf{1.000000} & \textbf{1.000000} & 80  & 3 \\
cfi-22  & \textbf{Laplacian} & \textbf{1.000000} & \textbf{1.000000} & 88  & 3 \\
cfi-24  & \textbf{Laplacian} & \textbf{1.000000} & \textbf{1.000000} & 96  & 3 \\
cfi-26  & \textbf{Laplacian} & \textbf{1.000000} & \textbf{1.000000} & 104 & 3 \\
cfi-28  & \textbf{Laplacian} & \textbf{1.000000} & \textbf{1.000000} & 112 & 3 \\
cfi-30  & \textbf{Laplacian} & \textbf{1.000000} & \textbf{1.000000} & 120 & 3 \\
cfi-32  & \textbf{Laplacian} & \textbf{1.000000} & \textbf{1.000000} & 128 & 3 \\
cfi-34  & \textbf{Laplacian} & \textbf{1.000000} & \textbf{1.000000} & 136 & 3 \\
\bottomrule
\end{longtable}

\subsection{rnd-3-reg Graphs}

\begin{table}[ht]
\centering
\caption{Results on random 3-regular graphs.}
\setlength{\tabcolsep}{2pt}
\begin{tabular}{lrrrrr c}
\toprule
Graph & $n$ & Method & NMI & $P$ & Classes & $k$-WL \\
\midrule
rnd-3-reg-1000-(1--11)  & 1000  & $G$-Laplacian & 1.000000 & 1.000000 & 1000 & $>$2 \\
rnd-3-reg-2000-(1--11)  & 2000  & $G$-Laplacian & 1.000000 & 1.000000 & 2000 & $>$2 \\
rnd-3-reg-3000-(1--11)  & 3000  & $G$-Laplacian & 1.000000 & 1.000000 & 3000 & -- \\
rnd-3-reg-4000-(1--11)  & 4000  & $G$-Laplacian & 1.000000 & 1.000000 & 4000 & -- \\
rnd-3-reg-5000-(1--11)  & 5000  & $G$-Laplacian & 1.000000 & 1.000000 & 5000 & -- \\
rnd-3-reg-6000-(1--11)  & 6000  & $G$-Laplacian & 1.000000 & 1.000000 & 6000 & -- \\
rnd-3-reg-7000-(1--11)  & 7000  & $G$-Laplacian & 1.000000 & 1.000000 & 7000 & -- \\
rnd-3-reg-8000-(1--11)  & 8000  & $G$-Laplacian & 1.000000 & 1.000000 & 8000 & -- \\
rnd-3-reg-9000-(1--11)  & 9000  & $G$-Laplacian & 1.000000 & 1.000000 & 9000 & -- \\
rnd-3-reg-10000-(1--11) & 10000 & $G$-Laplacian & 1.000000 & 1.000000 & 10000 & -- \\
\bottomrule
\end{tabular}
\end{table}

\subsection{KEF Graphs}

\begin{longtable}{l r c c c c c}
\caption{Results on KEF graphs.}
\label{tab:kef-results}\\

\toprule
\textbf{Graph} & \textbf{$n$} & \textbf{Method} &
\textbf{NMI} & \textbf{$P$} & \textbf{Classes} & \textbf{$k$-WL} \\
\midrule
\endfirsthead

\toprule
\textbf{Graph} & \textbf{$n$} & \textbf{Method} &
\textbf{NMI} & \textbf{$P$} & \textbf{Classes} & \textbf{$k$-WL} \\
\midrule
\endhead

kef10 & 200 & \textbf{$G$-Laplacian} & \textbf{1.000000} & \textbf{1.000000} & \textbf{19} & 3 \\
\midrule
kef11 & 242 & \textbf{$G$-Laplacian} & \textbf{1.000000} & \textbf{1.000000} & \textbf{23} & 3 \\
\midrule
kef12 & 288 & \textbf{$G$-Laplacian} & \textbf{1.000000} & \textbf{1.000000} & \textbf{24} & 3 \\
\midrule
kef13 & 338 & \textbf{$G$-Laplacian} & \textbf{1.000000} & \textbf{1.000000} & \textbf{27} & -- \\
\midrule
kef14 & 392 & \textbf{$G$-Laplacian} & \textbf{1.000000} & \textbf{1.000000} & \textbf{28} & -- \\
\midrule
kef15 & 450 & \textbf{$G$-Laplacian} & \textbf{1.000000} & \textbf{1.000000} & \textbf{31} & -- \\
\midrule
kef16 & 512 & \textbf{$G$-Laplacian} & \textbf{1.000000} & \textbf{1.000000} & \textbf{32} & -- \\
\midrule
kef17 & 578 & \textbf{$G$-Laplacian} & \textbf{1.000000} & \textbf{1.000000} & \textbf{35} & -- \\
\midrule
kef18 & 648 & \textbf{$G$-Laplacian} & \textbf{1.000000} & \textbf{1.000000} & \textbf{36} & -- \\
\midrule
kef19 & 722 & \textbf{$G$-Laplacian} & \textbf{1.000000} & \textbf{1.000000} & \textbf{39} & -- \\
\midrule
kef20 & 722 & \textbf{$G$-Laplacian} & \textbf{1.000000} & \textbf{1.000000} & \textbf{40} & -- \\
\bottomrule
\end{longtable}

\subsection{TRAN Graphs}

\small
\begin{longtable}{lrlcccc}
\caption{Results on TRAN graphs.}
\label{tab:tran-results}\\
\toprule
\textbf{Graph} & $\mathbf{n}$ & \textbf{Method} & \textbf{NMI} & $\mathbf{P}$ & \textbf{\# Classes} & \textbf{$k$-WL} \\
\midrule
\endfirsthead
\toprule
\textbf{Graph} & $\mathbf{n}$ & \textbf{Method} & \textbf{NMI} & $\mathbf{P}$ & \textbf{\# Classes} & \textbf{$k$-WL} \\
\midrule
\endhead
\endfoot
\bottomrule
\endlastfoot

tran\_6\_6.bliss & 6 & \textbf{$G$-Laplacian} & \textbf{1.000000} & \textbf{1.000000} & \textbf{1} & 1 \\
tran\_8\_12.bliss & 8 & \textbf{$G$-Laplacian} & \textbf{1.000000} & \textbf{1.000000} & \textbf{1} & 1 \\
tran\_10\_20.bliss & 10 & \textbf{$G$-Laplacian} & \textbf{1.000000} & \textbf{1.000000} & \textbf{1} & 1 \\
tran\_12\_24.bliss & 12 & \textbf{$G$-Laplacian} & \textbf{1.000000} & \textbf{1.000000} & \textbf{1} & 1 \\
tran\_16\_56.bliss & 16 & \textbf{$G$-Laplacian} & \textbf{1.000000} & \textbf{1.000000} & \textbf{1} & 1 \\
tran\_20\_120.bliss & 20 & \textbf{$G$-Laplacian} & \textbf{1.000000} & \textbf{1.000000} & \textbf{1} & 1 \\
tran\_24\_48.bliss & 24 & \textbf{$G$-Laplacian} & \textbf{1.000000} & \textbf{1.000000} & \textbf{1} & 1 \\
tran\_24\_72.bliss & 24 & \textbf{$G$-Laplacian} & \textbf{1.000000} & \textbf{1.000000} & \textbf{1} & 1 \\
\end{longtable}

\subsection{TNN Graphs}

\begin{longtable}{lrlcccc}
\caption{Results on TNN graphs.}
\label{tab:tnn-results}\\

\toprule
\textbf{Graph} & $\mathbf{n}$ & \textbf{Method} &
\textbf{NMI} & $\mathbf{P}$ & \textbf{\# Classes} &
\textbf{$k$-WL} \\
\midrule
\endfirsthead

\toprule
\textbf{Graph} & $\mathbf{n}$ & \textbf{Method} &
\textbf{NMI} & $\mathbf{P}$ & \textbf{\# Classes} &
\textbf{$k$-WL} \\
\midrule
\endhead

\bottomrule
\endfoot

tnn(1)\_26-1 & 26 & \textbf{$G$-Laplacian} & \textbf{1.000000} & \textbf{1.000000} & \textbf{12} & 3 \\
tnn(1)\_26-2 & 26 & \textbf{$G$-Laplacian} & \textbf{1.000000} & \textbf{1.000000} & \textbf{5} & 3 \\

tnn(2)\_52-1 & 52 & \textbf{$G$-Laplacian} & \textbf{1.000000} & \textbf{1.000000} & \textbf{12} & 3 \\
tnn(2)\_52-2 & 52 & \textbf{$G$-Laplacian} & \textbf{1.000000} & \textbf{1.000000} & \textbf{12} & 3 \\

tnn(3)\_78-1 & 78 & \textbf{$G$-Laplacian} & \textbf{1.000000} & \textbf{1.000000} & \textbf{12} & 3 \\
tnn(3)\_78-2 & 78 & \textbf{$G$-Laplacian} & \textbf{1.000000} & \textbf{1.000000} & \textbf{12} & 3 \\

tnn(4)\_104-1 & 104 & \textbf{$G$-Laplacian} & \textbf{1.000000} & \textbf{1.000000} & \textbf{12} & 3 \\
tnn(4)\_104-2 & 104 & \textbf{$G$-Laplacian} & \textbf{1.000000} & \textbf{1.000000} & \textbf{12} & 3 \\
\end{longtable}

\subsection{AG Graphs}

All 23 AG graphs were exactly resolved by the Laplacian of G at $r=10^{6}$, yielding $\mathrm{NMI}=1$, $P=1$, and two classes in every case. The vertex-orbit partition is also recovered by 1-WL for all these instances.

\subsection{Paley Graphs}

All 53 tested Paley graph instances were exactly resolved by the Laplacian of G, yielding $\mathrm{NMI}=1$, $P=1$, and a single equivalence class in every case. The vertex-orbit partition is also recovered by 1-WL for all these instances.

\subsection{PG Graphs}

All 23 tested PG graph instances were exactly resolved by the Laplacian of G at $r=10^{6}$, yielding $\mathrm{NMI}=1$, $P=1$, and a single equivalence class in every case. The vertex-orbit partition is also recovered by 1-WL for all these instances.

\subsection{Triang Graphs}

All 27 tested triangular graph instances were exactly resolved by the Laplacian of G at $r=10^{6}$, yielding $\mathrm{NMI}=1$, $P=1$, and a single equivalence class in every case. The vertex-orbit partition is also recovered by 1-WL for all these instances.

\subsection{K Graphs}

All 99 tested clique graph instances were exactly resolved by the Laplacian of G, yielding $\mathrm{NMI}=1$, $P=1$, and a single equivalence class in every case. The vertex-orbit partition is also recovered by 1-WL for all these instances.

\subsection{Latin Graphs}

All 29 tested Latin graph instances were exactly resolved by the Laplacian of G, yielding $\mathrm{NMI}=1$, $P=1$, and a single equivalence class in every case. The vertex-orbit partition is also recovered by 1-WL for all these instances.

\subsection{Lattice Graphs}

All 27 tested Lattice graph instances were exactly resolved by the Laplacian of G, yielding $\mathrm{NMI}=1$, $P=1$, and a single equivalence class in every case. The vertex-orbit partition is also recovered by 1-WL for all these instances.

\end{document}